\documentclass[aps,prb,reprint,twocolumn,superscriptaddress,nofootinbib]{revtex4-2}

\usepackage{graphicx}
\usepackage{bm}
\usepackage{amsmath}  
\usepackage{amssymb}
\usepackage{hyperref}
\usepackage{xcolor}
\usepackage{relsize}
\usepackage{array}
\usepackage{dsfont}
\usepackage{multirow}
\usepackage{physics}
\usepackage{lipsum}
\usepackage[T1]{fontenc}
\usepackage{booktabs}

\renewcommand{\v}[1]{{\boldsymbol{#1}}}
\newcommand{\ov}{\overline}

\newcommand{\bq}{{\v q}}

\begin{document}

\title{Fractional Chern mosaic in supermoir\'e graphene} 

\author{Yves H. Kwan}
\affiliation{Princeton Center for Theoretical Science, Princeton University, Princeton NJ 08544, USA}
\author{Tixuan Tan}
\affiliation{Department of Physics, Stanford University, Stanford, CA 94305, USA}
\author{Trithep Devakul}
\affiliation{Department of Physics, Stanford University, Stanford, CA 94305, USA}

\begin{abstract}
We propose the realization of a fractional Chern mosaic: a state characterized by a spatially varying topological order.
This state is enabled by a separation of length scales that emerges when three graphene sheets are sequentially rotated by a small twist angle.
The resulting structure features not only conventional moir\'e lattices, but also a much larger supermoir\'e lattice.
We demonstrate that a fractional Chern mosaic arises when electron correlations induce fractionalization locally on the moir\'e scale, while the pattern of fractionalization varies at the supermoir\'e scale.

\end{abstract}\maketitle

\section{Introduction}

{

The importance of length scales in physics is exemplified in the rise of two-dimensional moir\'e materials.
Twist engineered moir\'e lattices on the nanometer scale provide an ideal platform for experimentally realizing a remarkable variety of quantum phenomena~\cite{andrei2021marvels,mak2022semiconductor,kennes2021moire,wilson2021excitons}, from superconductivity~\cite{cao2018unconventional,yankowitz2019tuning,lu2019superconductors,chen2019signatures,arora2020superconductivity,park2021tunable,hao2021electric,xia2024unconventionalsuperconductivitytwistedbilayer,guo2024superconductivitytwistedbilayerwse2} to fractional Chern insulators~\cite{spanton2018observation,xie2021fractional,cai2023signatures,zeng2023thermodynamic,PhysRevX.13.031037,park2023observation,lu2024fractional}.
This rich physics is enabled by a separation of length scales between the atomic and moir\'e lattices.
At experimentally relevant scales, the physics is governed by that of the moir\'e lattice, which behaves as a new artificial crystal.

The interplay of length scales is at the forefront in an emerging class of 
 moir\'e materials with multiple incommensurate moir\'e lattices.
Prominent examples are various multilayer graphene-hBN heterostructures~\cite{lai2023imaging, wang2019composite, wang2019new, finney2019tunable, al2024topological, mao2021quasiperiodicity, leconte2020commensurate, andjelkovic2020double,shi2021moire,grover2022chern,rothstein2024band}
and twisted trilayer graphene at small, generic, twist angles~\cite{mora2019flatbands,zhu2020twisted,zhu2020modeling,zhang2021correlated,devakul2023magicangle,nakatsuji2023multiscale,guerci2023chern,mao2023supermoire,popov2023magic,popov2023magic2,foo2024extended,guerci2023nature,long2024evolution,park2024tunable,ren2023tunable,craig2024local,hao2024robust,uri2023superconductivity,yang2023multi,xia2023helicaltrilayergraphenemoire,kwan2024strong,dunbrack2023intrinsically}.
The presence of two incommensurate moir\'e lattices introduces an additional large supermoir\'e (or ``moir\'e-of-moir\'e'') length scale, which can be parametrically larger than the moir\'e lattices. 
This new supermoir\'e scale leads to another separation of length and energy scales, opening the door to especially intriguing correlated electronic states.

One especially striking example is the Chern mosaic \cite{Grover_2022,
PhysRevB.103.075122,PhysRevResearch.6.L022025,PhysRevB.103.075423,kwan2021domain,PhysRevLett.128.156801}.
In this state, electrons form Chern insulators at the moir\'e scale, but the Chern number varies across the supermoir\'e scale.
The result is a mosaic of Chern insulators embedded in a network of topological domain walls.
Here, we consider an even more exotic possibility: a ``fractional Chern mosaic'', where electrons spontaneously fractionalize within local regions of fractional Chern insulators (FCI)~\cite{parameswaranFractionalQuantumHall2013,bergholtzTOPOLOGICALFLATBAND2013,liuRecentDevelopmentsFractional2022,neupertFractionalQuantumHall2011,shengFractionalQuantumHall2011,regnaultFractionalChernInsulator2011,sun2011nearly}, but with the pattern of fractionalization varying at the supermoir\'e scale.
Such a fractional Chern mosaic would represent a novel class of quantum states characterized by a spatially varying topological order.

We propose the realization of a fractional Chern mosaic.
Specifically, we focus on helical trilayer graphene (HTG), a prototypical supermoir\'e material consisting of three graphene sheets sequentially rotated by a small twist angle $\theta$, i.e. in the configuration $(\theta_1,\theta_2,\theta_3)=(\theta,0,-\theta)$.  
In HTG, adjacent layer pairs produce moir\'e lattices with periods $a_m\approx a_0/\theta$, which combine to form a large supermoir\'e lattice with period $a_{sm}\approx a_0/\theta^2$, where $a_0=0.246$nm is the graphene lattice constant.
This supermoir\'e lattice, on the scale of a few hundred nanometers, was recently imaged with a scanning single-electron transistor (SET)~\cite{hoke2024imagingsupermoirerelaxationconductive}.
Lattice relaxation creates large triangular moir\'e-periodic domains, referred to as h-HTG and $\overline{\mathrm{h}}$-HTG, as illustrated in Fig.~\ref{fig:nonint_DOS}~\cite{devakul2023magicangle,nakatsuji2023multiscale}.
Due to the separation of lengthscales $a_{sm}\gg a_m$, it suffices to model the system as locally periodic within each domain.
Within these periodic domains, the local moir\'e band structure gives rise to topological flat bands with valley-contrasting Chern number and opposing topology in adjacent domains~\cite{
devakul2023magicangle,nakatsuji2023multiscale,guerci2023chern}.
Flat band ferromagnetism then leads to spontaneous time reversal symmetry breaking, as observed near a magic angle~$\theta\approx1.8^\circ$~\cite{xia2023helicaltrilayergraphenemoire}. 
The combined presence of broken time-reversal symmetry, flat Chern bands, and supermoir\'e periodicity sets the stage for the fractional Chern mosaic.  

The task is to establish whether electrons within each domain can fractionalize.
To achieve this, we perform a comprehensive analysis of the correlated physics at partial filling of the flat Chern bands.  
We begin by characterizing an additional momentum-dependent tunneling (MDT) term in the Hamiltonian,
which originates from the decay of the interlayer tunneling amplitude as a function of the in-plane distance between the atomic $p_z$ orbitals~\cite{koshino2015interlayer}.
We find that accounting for MDT not only explains the particle-hole asymmetry observed in transport~\cite{xia2023helicaltrilayergraphenemoire} but,
crucially,
 stabilizes an FCI at filling factor $\nu=3+1/3$.
The predicted state, illustrated in Fig.~\ref{fig:nonint_DOS}a, features electron fractionalization into charge $e/3$ quasiparticles in a spatially varying way due to the opposing band topology in the h-HTG and $\overline{\mathrm{h}}$-HTG domains, and can be experimentally unveiled by local scanning probes.

\begin{figure}[t!]
    \centering
    \includegraphics[width=1\linewidth]{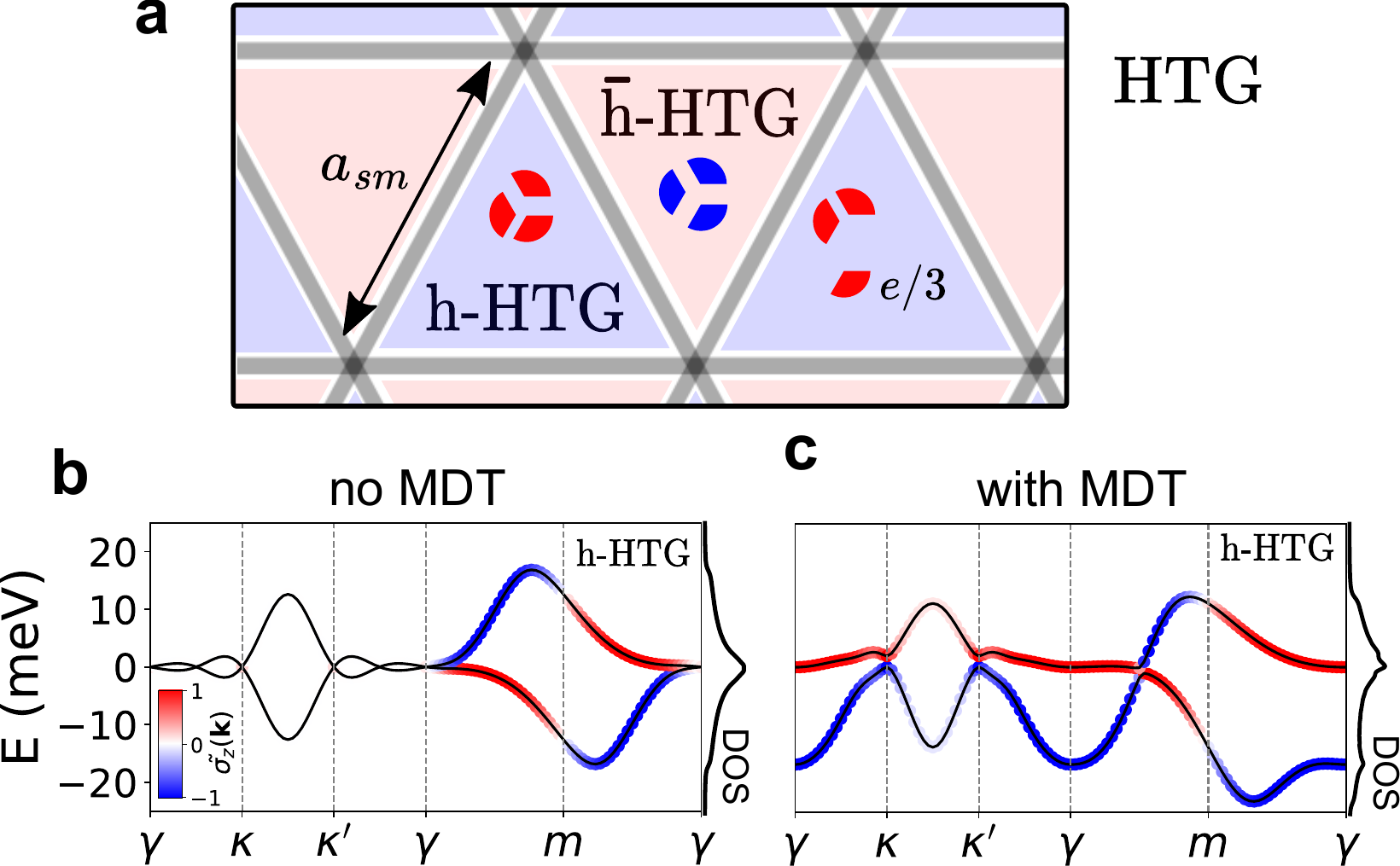}
    \caption{\textbf{a)} Helical trilayer (HTG) relaxes on the supermoir\'e scale $a_{sm}$ into h-HTG and $\bar{\text{h}}$-HTG domains that are locally moir\'e-periodic, and separated by gapless domain walls (grey). Fractionalization within the domains generates a fractional Chern mosaic. \textbf{b)} Band structure in valley $K$ of the h-HTG domain in the absence of momentum-dependent tunneling (MDT). The central two bands are colored according to Chern-sublattice polarization $\tilde{\sigma}_z(\bm{k})$, where red (blue) indicates polarization onto the $A$ ($B$) sublattice. Right line trace plots the density of states. \textbf{c)} Same as b) except in the presence of MDT.}
    \label{fig:nonint_DOS}
\end{figure}

}

\textit{Non-interacting model}---Our starting point is a generalized Bistritzer-MacDonald~\cite{bistritzerMoireBandsTwisted2011} continuum model description of the local band structure within each h-HTG (or $\bar{\text{h}}$-HTG) domain~\cite{devakul2023magicangle}.
The $K$ valley of h-HTG is described by
\begin{equation}\label{eq:nonint_BM}
H_K=\begin{bmatrix}
    -iv\bm{\sigma}\cdot\nabla & T(\bm{r}-\bm{d}_t) & 0 \\
    T^\dagger(\bm{r}-\bm{d}_t) & -iv\bm{\sigma}\cdot\nabla & T(\bm{r}-\bm{d}_b)\\
    0 & T^\dagger(\bm{r}-\bm{d}_b) & -iv\bm{\sigma}\cdot\nabla
\end{bmatrix}
\end{equation}
where $\bm{\sigma}=(\sigma_x,\sigma_y)$ acts on the microscopic sublattice, and $v=8.8\times 10^5\,\text{ms}^{-1}$. 

In the absence of MDT, the interlayer tunneling is
\begin{equation}\label{eq:nonint_BM_tunnel}
\begin{gathered}
    T(\bm{r})=\begin{bmatrix}
        w_{AA}t_0(\bm{r}) & w_{AB}t_{-1}(\bm{r})\\
        w_{AB}t_1(\bm{r}) & w_{AA}t_0(\bm{r})
    \end{bmatrix}\\
    t_\alpha(\bm{r})=\sum_{n=0}^{2}e^{\frac{2\pi i}{3}n\alpha}e^{i\bm{q}_n\cdot\bm{r}}\\
    q_{n,x}+iq_{n,y}=-ik_\theta e^{\frac{2\pi i}{3}n},
\end{gathered}
\end{equation}
where $k_\theta=2K_D\sin\frac{\theta}{2}$, with $K_D=\frac{4\pi}{3a_0}$ the Dirac wavevector, and we fix $(w_{AA},w_{AB})=(75\,\text{meV},110\,\text{meV}$). 
Within the h-HTG domain, the relative interlayer moir\'e shift is $\bm{d}_t-\bm{d}_b=\frac{1}{3}(\bm{a}_2-\bm{a}_1)$, where $\bm{a}_{1,2}=\frac{4\pi}{3k_\theta}(\pm\frac{\sqrt{3}}{2},\frac{1}{2})$ are the basis moir\'e vectors.
We focus only on the physics within an h-HTG domain, as the $\bar{\text{h}}$-HTG domains are related by $\hat{C}_{2z}$ symmetry.

Eq.~\ref{eq:nonint_BM} and \ref{eq:nonint_BM_tunnel} obey an exact particle-hole-inversion symmetry\footnote{This is only weakly broken when introducing the small twists in the Pauli terms.} (PHS) that is strongly broken in experiment~\cite{xia2023helicaltrilayergraphenemoire}, where correlated states appear only for electron doping.
This symmetry is broken by considering the finite range of interlayer hopping, which generates MDT.
This is incorporated to leading order by making the replacement
\begin{equation}
t_{\alpha}(\bm{r}) \rightarrow \sum_{n=0}^{2} e^{\frac{2\pi i}{3}n\alpha} e^{i\bm{q}_{n}\cdot\bm{r}}(1 + \lambda_{\mathrm{MDT}} \hat{\bm{q}}_{n,\perp}\cdot \vec{\nabla}),
\end{equation}
where $\hat{\bm{q}}_{n,\perp}= \left[\cos(\frac{2\pi}{3}n),\sin(\frac{2\pi}{3}n)\right]$ is a unit vector perpendicular to $\bm{q}_n$. 
We take a realistic $\lambda_{\text{MDT}}\approx -2.3\,$\r{A}, as derived in App.~\ref{secapp:kdt}. 

Fig.~\ref{fig:nonint_DOS}b,c shows the non-interacting band structure of $\theta=1.8^\circ$ h-HTG within a single spin-valley sector. Regardless of MDT, we obtain a pair of narrow bands separated by large gaps $\gtrsim 80\,$meV to higher remote bands that are not shown. 
The non-trivial topology is unveiled by diagonalizing the sublattice operator $\sigma_z$ projected to the central bands~\cite{bultinckGroundStateHidden2020,lianTBGIVExact2020,ledwithStrongCouplingTheory2021}. This yields the so-called Chern basis indexed by a `Chern-sublattice' label $\tilde{\sigma}=A,B$ according to the dominant sublattice polarization\footnote{$\tilde{\sigma}$ refers to the Chern basis, while $\sigma$ indexes the microscopic sublattice. They become equivalent in the chiral limit. We will often use the terms Chern and sublattice interchangably.}. Within valley $K$, these bands have Chern numbers
\begin{equation}\label{eq:strongcoupling_Chern}
    C_{K,A}=1,\quad C_{K,B}=-2,
\end{equation}
and the corresponding values for valley $\bar{K}$ and the $\bar{\text{h}}$-HTG domain can be obtained by time-reversal and $\hat{C}_{2z}$. 
The blue and red colors indicate the expectation value of $\langle \tilde{\sigma}_z \rangle\equiv \langle P_A\rangle-\langle P_B\rangle $, where $P_{\tilde{\sigma}}$ is the projector onto the $\tilde{\sigma}$ band.

In the absence of MDT (Fig.~\ref{fig:nonint_DOS}b), the central bands obey PHS as reflected in the density of states. 
Band eigenstates with predominantly $B$ character $(\langle\tilde{\sigma}_z\rangle\approx-1)$ are more dispersive than those with $A$ character, but both have the same energy when averaged over the mBZ. To make this concrete, we can rotate to the Chern basis and switch off the off-diagonal hybridization to define the dispersions $E_{\tilde{\sigma}_z}(\bm{k})$  (see App.~\ref{secapp:interChern}). At $\theta=1.8^\circ$, we find that $E_A(\bm{k})$ and $E_B(\bm{k})$ have zero average, but bandwidths $7.2\,$meV and $30.4\,$meV respectively.

In contrast, introduction of MDT (Fig.~\ref{fig:nonint_DOS}c) lowers the energy of the $B$ basis relative to the $A$ basis. At $\theta=1.8^\circ$, $E_A(\bm{k})$ has a mean energy of $1.0\,$meV and is bounded by $-2.9\,$meV and $4.2\,$meV, while $E_B(\bm{k})$ has a mean energy of $-6.4\,$meV and is bounded by $-22.4\,$meV and $9.4\,$meV. This strongly breaks the PHS, as evidenced by the substantial asymmetry of the density of states. 
These details will become important soon when considering correlated states.

\textit{Interactions}---We consider gate-screened 
Coulomb interactions $V(q)=\frac{e^2}{2\epsilon_0\epsilon_r q}\tanh qd_{\text{sc}}$, where the gate distance is $d_\text{sc}=25\,\text{nm}$ and the relative permittivity $\epsilon_r\simeq 8-15$ is left as a tuning parameter. The large single-particle remote gaps permit projection onto the two central bands per spin and valley
\begin{equation}
H_\text{int} = \frac{1}{2A} \sum_{\bq } V(\bq) \delta \hat{\rho}_\bq \delta \hat{\rho}_{-\bq}
\label{eq:intham_main}
\end{equation}
where $A$ is the system area, and  $\delta \hat{\rho}_\bq = \hat{\rho}_{\bq} - 4 \ov{\rho}_\bq$ is the projected density operator measured relative to the mean density $\ov{\rho}_\bq$ of the central bands (this is the so-called `average' interaction scheme)~\cite{kwan2024strong}. $H_\text{int}$ satisfies a $U(2)_K\times U(2)_{\bar{K}}$ flavor symmetry, which includes charge-$U(1)_c$ and valley-$U(1)_v$ conservation, and valley-dependent $SU(2)_s$ spin-rotations.

\begin{figure*}[t!]
    \centering
    \includegraphics[width=0.9\linewidth]{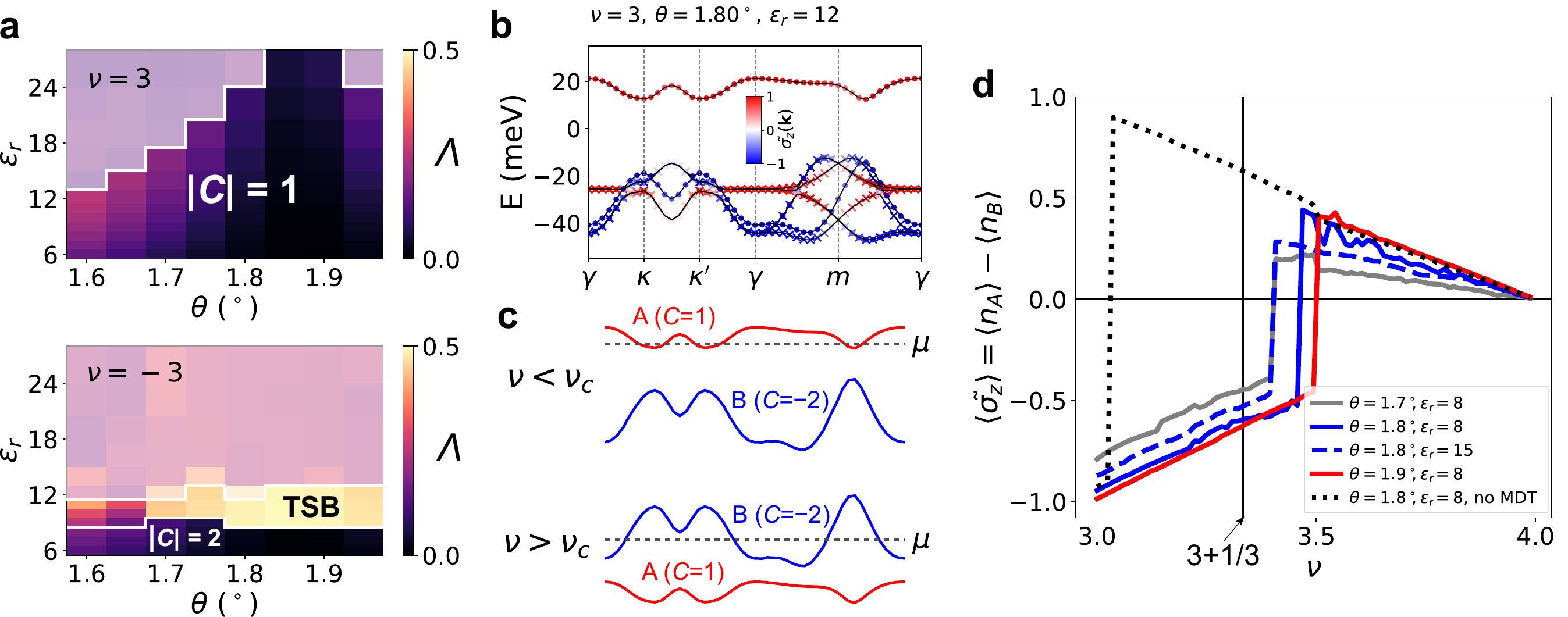}
    \caption{\textbf{a)} HF phase diagram as a function of twist angle $\theta$ and relative permittivity $\epsilon_r$ at $\nu=+3$ (top) and $\nu=-3$ (bottom). White shaded regions indicate gapless states. $\Lambda$, defined in the main text, 
    is a measure of how far the HF solution is from an idealized strong-coupling state. 
    TSB indicates various states that break moir\'e translation symmetry. System size is $12\times 12$, and we allow for spontaneous doubling or tripling of the moir\'e unit cell along both directions. \textbf{b)} HF band structure of the $\nu=+3$ Chern insulator. The bands are colored according to their Chern-sublattice polarization. Dots (crosses) denote spin up (down). System size is $24\times 24$. \textbf{c)} Schematic of Chern-sublattice transitions induced by the sublattice asymmetry in the effective dispersion. 
    \textbf{d)} HF calculations of Chern-sublattice polarization $\langle \tilde{\sigma}_z\rangle$ for fillings $3\leq\nu\leq 4$. $\langle \tilde{\sigma}_z\rangle$ is defined as the difference of the occupation numbers of the $A$ and $B$ Chern-sublattice bands per moir\'e unit cell. We enforce moir\'e translation symmetry and valley $U_v(1)$ symmetry. Calculations are performed with MDT unless otherwise indicated. System size is $18\times 18$.}
    
    \label{fig:HF_figure}
\end{figure*}

\textit{Hartree-Fock phase diagram}---
We first consider the physics at integer fillings within Hartree-Fock (HF) theory.
In the absence of MDT, Ref.~\cite{kwan2024strong}  identified a hierarchy of so-called ``strong-coupling'' insulators at integer fillings $\nu$. These phases can be understood by fully polarizing into $\nu+4$ of the eight total Chern bands indexed by spin $s\in\{\uparrow,\downarrow\}$, valley $\tau\in \{K,K^\prime\}$, and Chern-sublattice $\tilde{\sigma}_z\in\{A,B\}$, with the associated Chern number following from Eq.~\ref{eq:strongcoupling_Chern}. 
The phase diagram in Ref.~\cite{kwan2024strong} was identical under $\nu\leftrightarrow -\nu$ due to PHS.   

In contrast, MDT leads to qualitatively different behaviors at negative and positive integer fillings in our HF calculations (Fig.~\ref{fig:HF_figure}a). We diagnose the strong-coupling nature by the deviation from full ($\tau,s,\tilde{\sigma}$)-polarization 
\begin{equation}
\Lambda\equiv \text{max}_{\tau,s,\tilde{\sigma}}\Big[\text{min}\big(\langle n_{\tau,s,\tilde{\sigma}}\rangle,1-\langle n_{\tau,s,\tilde{\sigma}}\rangle\big)\Big],
\end{equation}
where $\langle n_{\tau,s,\tilde{\sigma}}\rangle$ is the mBZ-averaged occupation number for the Chern band $\tilde{\sigma}$ in valley $\tau$ and spin $s$. $\Lambda$ vanishes for an idealized strong-coupling insulator, while it can take values as large as $0.5$ for generic states. For $\nu=+3$, we find a strong-coupling $|C|=1$ insulator for a range of $\theta$, including for very weak interactions $\epsilon_r\gtrsim 25$. 
On the other hand for $\nu=-3$, the strong-coupling phase only exists for for relatively strong interactions $\epsilon_r\lesssim 9$.
Lowering the interaction strength leads to gapless states, or various gapped phases with intervalley coherence and translation symmetry-breaking~\cite{kwan2024strong,wang2024cherntexture}. Analogous results hold for other integer fillings (Fig.~\ref{figapp:HF_phase_diagram_all_int}), demonstrating that MDT can explain the pronounced particle-hole asymmetry observed in transport~\cite{xia2023helicaltrilayergraphenemoire}.

Fig.~\ref{fig:HF_figure}b shows the HF band structure of the $|C|=1$ insulator at $\nu=+3$. The state is fully spin-valley polarized within the flat-band manifold, and the strong-coupling nature is evidenced by the near total $\tilde{\sigma}_z$ polarization of the conduction band. In the partially-filled flavor, the system occupies the $B$ band (blue) with lower mean kinetic energy, leaving a narrow unfilled $A$ band (red) with $|C|=1$.

\textit{Sublattice transitions}---At $\nu=3$, the unfilled conduction band is a narrow $|C|=1$ band of predominantly $\tilde{\sigma}=A$ character, with favorable quantum geometry (Fig.~\ref{figapp:HF_QG}) for hosting FCIs when fractionally occupied. 
For this to occur, it is crucial that the notion of an isolated conduction $A$ band remains valid at higher fractional fillings like $\nu=3+\frac{1}{3}$.

To this end, we discover a cascade of sublattice transitions (Fig.~\ref{fig:HF_figure}c,d), whose mechanism relies on two key ingredients: (i) sublattice ferromagnetism induced by the dominant Coulomb interactions, and (ii) the lower average but higher peak kinetic energy of the $B$ band compared to the $A$ band. In the following, we discuss the cascade for the filling range $3\leq \nu \leq 4$ assuming full flavor polarization (i.e. flat bands in three of four spin-valley flavors are fully filled). Exactly at $\nu=3$, occupying either Chern band of the remaining flavor yields a valid exchange-driven strong-coupling insulator, but the system chooses to form the $|C|=1$ insulator and fill the $B$ band due to its lower average kinetic energy, yielding a $\langle\tilde{\sigma}_z\rangle\approx-1$ state. As electrons are doped into the system, they naturally populate the empty $A$ band (Fig.~\ref{fig:HF_figure}c, top), which gradually degrades the initial negative sublattice polarization $-1<\langle \tilde{\sigma}_z\rangle<0$. Eventually we reach a critical filling $\nu_c$ where the system energetically prefers to instead fully occupy the $A$ band and partially occupy the $B$ band  so that $\langle \tilde{\sigma}_z\rangle>0$ (Fig.~\ref{fig:HF_figure}c, bottom). The reason is that doing so allows the system to empty the parts of the $B$ band that are kinetically costly, while respecting the tendency towards sublattice polarization. $\nu_c$ is determined by the interplay between the relative energy gain from fully polarizing the $B$ band over the $A$ band, and the kinetic cost of occupying the high energy parts of the $B$ band. The outcome is an abrupt sign change of the sublattice polarization at $\nu_c$.

Fig.~\ref{fig:HF_figure}d tracks the self-consistent HF sublattice polarization. For all parameters, $\langle \tilde{\sigma}_z\rangle$ starts off near $-1$ and increases linearly with a gradient close to 1 as the empty $A$ band is progressively occupied. With MDT, we find a strongly first-order sublattice jump at which $\langle \tilde{\sigma}_z\rangle$ changes sign. Crucially, the transition filling $\nu_c\approx 3.5$ is greater than $3+\frac{1}{3}$ for realistic parameters. The enhanced dispersion of the $B$ band means that the Fermi surfaces for $\nu>\nu_c$ are often flavor-degenerate, but the phase for $\nu<\nu_c$ is always fully flavor polarized. These findings are consistent with the experimental observation of AHE persisting to $\nu\approx 3.5$~\cite{xia2023helicaltrilayergraphenemoire}. 
$\nu_c$ is lowered when increasing $\epsilon_r$.
If MDT is neglected (dotted lines), the sublattice transition occurs almost immediately upon electron doping $\nu=+3$. The reason is that without MDT, the average kinetic energies of the $A$ and $B$ bands are identical (Fig.~\ref{fig:HF_figure}a), and only the sublattice exchange anisotropy of $\lesssim 0.5\,$meV favors occupying the $B$ band at $\nu=+3$~\cite{kwan2024strong}. This tiny splitting is easily swamped at small $\nu_c$ by the kinetically-driven mechanism explained above, underscoring the importance of MDT in faithfully capturing the sublattice physics.
Similar sublattice transitions occur between every non-negative integer filling, but none occur for $\nu<0$, where there is significantly less symmetry-breaking (Fig.~\ref{figapp:allnu}).

\begin{figure}[t!]
    \centering\includegraphics[width=\linewidth]{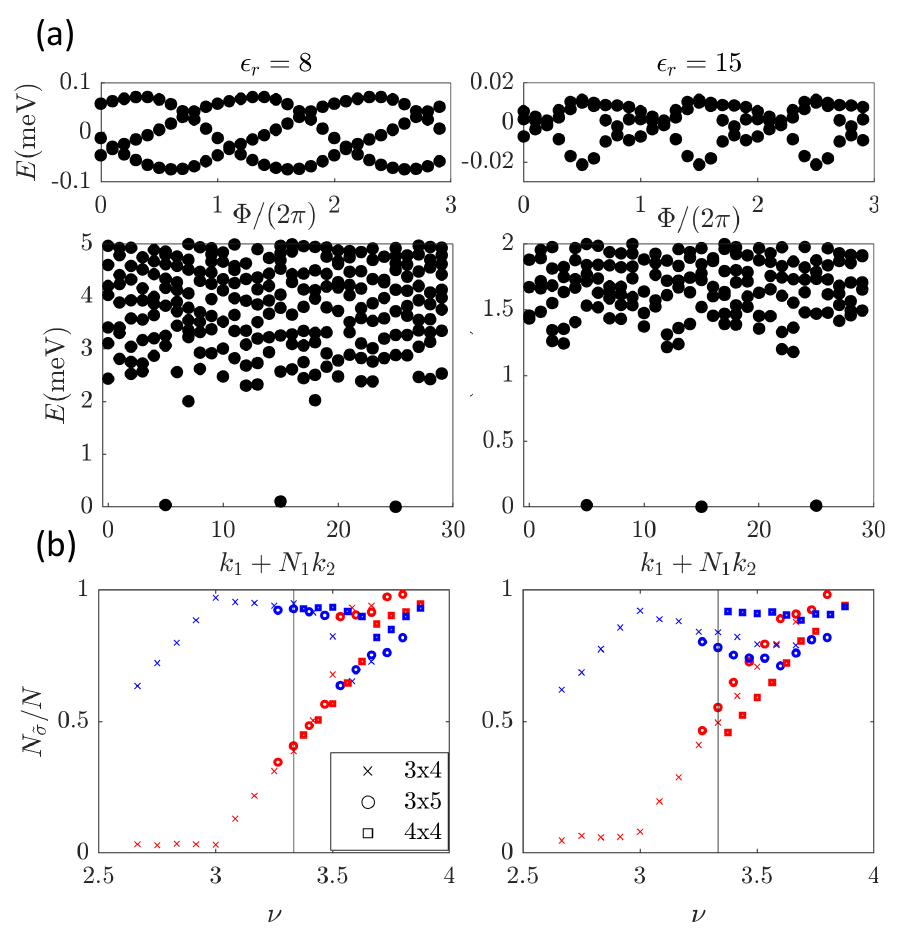}
    \caption{\textbf{a)} 1-band ED calculation on the $5\times6$ cluster at $\nu=3+1/3$. Bottom row shows the momentum-resolved many-body spectrum for $\epsilon_r=8$ and $15$. Top row shows the corresponding evolutions of the three quasi-degenerate ground states under flux threading $\Phi$. \textbf{b)} 2-band ED calculation of the Chern-sublattice polarization as a function of filling. Different symbols (see legend) refer to various lattice geometries with corresponding system sizes $N=12,15,16$. $N_{\tilde{\sigma}}$ is the expectation value of the particle occupation in the Chern-sublattice basis. Red (blue) corresponds to the $A$ ($B$) basis. The vertical line indicates $\nu=3+1/3$. All calculations are performed at $\theta=1.8^\circ$ and with MDT.}
    \label{fig:main_ED}
\end{figure}

\textit{Exact diagonalization}---We use exact diagonalization (ED) to investigate whether FCIs can be stabilized at fractional fillings. Since full calculations in the eight central bands are expensive, it is necessary to truncate the `active' part of the many-body Hilbert space. In Fig.~\ref{fig:main_ED}a, we first present 1-band ED results at $\nu=3+\frac{1}{3}$, where the active subspace is restricted to the $|C|=1$ conduction band of the $\nu=+3$ HF insulator. In other words, we constrain the other seven HF bands to be fully occupied\footnote{We therefore include the Hartree and Fock potentials from these non-active bands.}, and work at $1/3$ filling of the remaining Chern band which is spin-valley polarized and of predominantly $\tilde{\sigma}=A$ character (see Fig.~\ref{fig:HF_figure}b). 
We obtain an FCI for $\epsilon_r=8,15$ on a system with $N=30$ unit cells, as evident from the three-fold quasi-degenerate ground states with correct momenta that remain gapped under flux threading. In App.~\ref{secapp:addtional_ED}, we show $N=27$ results that rule out charge density waves.

Next, we perform ED calculations where we only constrain that three flavors  are fully occupied, leading to the restriction $2\leq \nu \leq 4$. We emphasize that this effective 2-band calculation is \emph{exact} within the maximally spin-valley-polarized symmetry sector of the eight central bands. From the HF results, we anticipate that the ground state indeed lies in this flavor sector from $\nu=+3$ up to at least the sublattice transition at $\nu_c$ (see App.~\ref{subsecapp:HF_noninteger}). Importantly, this 2-band calculation faithfully captures the interplay between the $\tilde{\sigma}=A,B$ bands. 

Fig.~\ref{fig:main_ED}b plots the Chern-sublattice densities $\langle n_A\rangle$ and $\langle n_B\rangle$  in the 2-band calculation as a function of $\nu$. 
Owing to the increased computational difficulty, we collate results for several system sizes. Consistent with the HF analysis, the $\nu=+3$ ground state is strongly polarized into the $B$ band and is essentially a gapped single Slater determinant (see App.~\ref{secapp:addtional_ED}). 
Upon electron-doping for $\epsilon_r=8$, the $B$ band maintains a large occupancy $>90\%$ such that the additional carriers primarily populate the $A$ band. Crucially, this picture still holds at $\nu=3+\frac{1}{3}$ across multiple lattice geometries, lending credence to the 1-band ED. The theory of sublattice transitions is corroborated by the abrupt jump at $\nu_c\approx3.5$. 
For $\epsilon_r=15$, the $B$ band depopulates more appreciably for certain system sizes when adding electrons to the $\nu=+3$ insulator, though we still find kinks in the sublattice occupations around $\nu_c\approx 3.5$. 
We note that direct signatures of FCIs in our multi-band calculations are present but not fully robust (see App.~\ref{secapp:2band_bandmixing}), likely due to finite-size effects since the largest size\footnote{The Hilbert space dimension per momentum at $N=18,21$ is approximately $7\times 10^7$ and $3\times 10^9$ respectively. }
for which we have 2-band data at $\nu=3+\frac{1}{3}$ is smaller than the threshold $N$ beyond which 1-band ED robustly yields FCIs.

\textit{Fractional Chern mosaics} ---
Having found evidence for FCIs within the h-HTG domains, we now zoom out and re-examine the structure of HTG at $\nu=3+1/3$ as a whole.
In the h-HTG domain, electrons within the $A$ band of a particular spin-valley ($\tau,s$) flavor fractionalizes. 
The resulting state has a local Hall conductivity
$\sigma_{xy}=-\frac{2}{3}\tau_z\frac{e^2}{h}$ where $\tau_z=1 $ (for $K$) or $-1$ (for $K^\prime$).
The fractional Hall conductivity can be decomposed as $-\frac{2}{3}=-1+\frac{1}{3}$, where $-1$ arises from the Chern insulator at $\nu=3$, and $1/3$ from the FCI.
Since the local physics in the $\bar{\text{h}}$-HTG domain is related by $\hat{C}_{2z}$, the corresponding FCI there will arise from a partially filled $B$ band
with local $\sigma_{xy}=\frac{2}{3}\tau_z \frac{e^2}{h}$.
Since the spin-valley polarization is determined spontaneously, there are multiple qualitatively distinct fractional Chern mosaics that may occur.
Which of these scenarios is favored will likely depend on fine details of the domain wall energetics~\cite{kwan2021domain,kwan2024strong}.

Consider the case where the valley polarizations in the h-HTG and $\bar{\text{h}}$-HTG domains are identical.
In this case, a mosaic of local $\sigma_{xy}=\pm\frac{2}{3}\frac{e^2}{h}$ domains is formed.  
The domain walls necessarily host fractionally charged topological chiral edge modes.
In both types of domains, electrons fractionalize into charge $e/3$ quasiparticles as is known for the $1/3$ Laughlin state.
However, the anyonic exchange phase $e^{\pm i\frac{\pi}{3}}$ will be opposite for quasiparticles in the two types of domains.
The result is a novel state in which fractionally charged quasiparticles exhibit spatially varying anyonic statistics. 

Another possibility is that the valley polarizations will be opposite in the two domains.
This configuration will be favored by a small external magnetic field.
For concreteness, suppose that the FCI arises in the $K$ valley for h-HTG, and in the $K^\prime$ valley for $\bar{\text{h}}$-HTG.
In this case, while both domains have local $\sigma_{xy}=-\frac{2}{3}\frac{e^2}{h}$, we consider this state a fractional Chern mosaic since the fractionally charged $e/3$ quasiparticles carry opposite valley quantum numbers in the two domains.

\textit{Discussion}---Our results have important implications for recent and future experiments.
The transport study of Ref.~\cite{xia2023helicaltrilayergraphenemoire} observed correlated features only for positive fillings in multiple magic-angle ($\theta\approx 1.7^\circ-1.8^\circ$) HTG devices, consistent with our HF calculations in the presence of MDT. 
The anomalous Hall effect and density-tuned hysteresis observed near $3\lesssim \nu\lesssim 3.5$ are consistent with the sublattice transition identified here via the kinetic mechanism at $\nu_c\approx 3.5$.
Combined with the recent observation of supermoir\'e domains in (non magic-angle) HTG by Ref.~\cite{hoke2024imagingsupermoirerelaxationconductive},
our analysis implies that a fractional Chern mosaic can be realized at $\nu=3+1/3$, but is not likely visible in transport~\cite{xia2023helicaltrilayergraphenemoire} due to the network of conducting domain walls and sample inhomogeneity.
Local probes are ideal for resolving the real-space mosaic and detecting the local topological state. 
Scanning single-electron transistors (SET)~\cite{xieFractionalChernInsulators2021,yu2022correlated,yu2022spin,hoke2024imagingsupermoirerelaxationconductive} can identify the state as an incompressible feature that disperses with perpendicular magnetic field $B$ according to the Streda formula. 
Scanning tunneling microscopy (STM)~\cite{wongCascadeElectronicTransitions2020,choi2021correlation,choi2021interactiondriven,nuckolls2023quantum,kim2023imaging,turkel2022orderly,nuckollsStronglyCorrelatedChern2020,zhang2023local} would further be able to directly image the cascade of sublattice transitions predicted in this work. 
We also note that the global $\hat{C}_{2z}$ symmetry of pristine HTG implies that identical physics would be present in the h-HTG and $\bar{\text{h}}$-HTG domains, but alignment to the hBN substrate could introduce a sublattice splitting that favors fractionalization in only one type of domain.

Finally, while we have mainly focused on $\nu=3+\frac{1}{3}$, 
FCIs may exist at other fillings, especially those which are compatible with the picture above. The simplest candidates are Jain states at $\nu=3+\frac{k}{2k+1}$ for which $\nu<\nu_c$.
The periodic resets in flavor and sublattice physics suggest that fractionalization could also occur around other integer fillings, a possibility which we leave for future theoretical studies.

In summary, we have predicted the realization of a fractional Chern mosaic enabled by a combination of local topological flat bands and supermoir\'e periodicity.
Our results present an opportunity to engineer and explore novel intricate quantum states with spatially varying topological order.

\begin{acknowledgments}

We thank Liqiao Xia, Aviram Uri, Ben Feldman, Sergio de la Barrera, Aaron Sharpe, and Jonah Herzog-Arbeitman for valuable discussions and comments. 
TD thanks Junkai Dong, Daniel Parker, Patrick Ledwith, Tomohiro Soejima, Ashvin Vishwanath, and Michael Zaletel for early contributions and discussions that inspired this project.
TD and TT are supported by a startup fund at Stanford University.
Some of the ED calculations were performed using the DiagHam software package.

\end{acknowledgments}

%

\newpage
\clearpage

\setcounter{section}{0}
\setcounter{figure}{0}
\let\oldthefigure\thefigure
\renewcommand{\thefigure}{S\oldthefigure}

\setcounter{table}{0}
\renewcommand{\thetable}{S\arabic{table}}

\renewcommand{\thesection}{S\arabic{section}}
\renewcommand{\thesubsection}{\thesection.\arabic{subsection}}
\renewcommand{\thesubsubsection}{\thesubsection.\arabic{subsubsection}}

\onecolumngrid

\begin{appendix}

\section{Derivation of momentum dependent tunneling term}\label{secapp:kdt}

In this Appendix, we give a derivation of the momentum dependent tunneling (MDT) term used in the main text.
For simplicity, consider twisted bilayer graphene with a small twist angle $\theta$.
The generalization to trilayer is straightforward.

Each graphene layer is described by a honeycomb lattice tight binding model with the two sublattices labeled $\sigma\in\{A,B\}$.
We keep spin labels implicit.
Working in momentum space, states in the layer $\ell=1,2$ are labeled by 
\begin{equation}
\ket{\bm{k},\sigma,\ell} = \frac{1}{\sqrt{N}}\sum_{\bm{R}_{\sigma \ell}} e^{i\bm{k}\cdot \bm{R}_{\sigma \ell}}\ket{\bm{R}_{\sigma \ell},\sigma,\ell}
\end{equation}
where $\bm{R}_{\sigma \ell}$ are the locations of the sublattice $\sigma$ sites on layer $\ell$,
$\ket{\bm{R}_{\sigma \ell},\sigma,\ell}$ is the corresponding state,
and $N$ is the total number of unit cells.
We assume the system is large but finite in order to properly normalize the wavefunctions.
Here, $\bm{k}$ is a crystal momentum defined within the monolayer Brillouin zone of layer $\ell$, so $\ket{\bm{k}+\bm{G}_\ell,\sigma,\ell}\equiv \ket{\bm{k},\sigma,\ell}$ where $\bm{G}_\ell$ is a monolayer reciprocal lattice vector of layer $\ell$.
The inverse Fourier transform is
\begin{equation}
\ket{\bm{R}_{\sigma \ell},\sigma,\ell} = \frac{1}{\sqrt{N}}\sum_{\bm{k}} e^{-i\bm{k}\cdot \bm{R}_{\sigma \ell}}\ket{\bm{k},\sigma,\ell}.
\end{equation}
where the sum is over momenta $\bm{k}$ within the appropriate Brillouin zone.
The intralayer Hamiltonian acts within each layer and gives rise to the usual monolayer graphene dispersions.

We are concerned with the interlayer tunneling.
Consider a general tunneling matrix element that depends only on the distance between the initial and final states (the two-center approximation),
\begin{equation}
H_{\text{inter}} = \sum_{\sigma^\prime \sigma} \sum_{\bm{R}_{\sigma 1} \bm{R}_{\sigma 2}} W(\bm{R}_{\sigma^\prime 2}-\bm{R}_{\sigma 1})\ket{\bm{R}_{\sigma^\prime 2},\sigma^\prime,2} \bra{\bm{R}_{\sigma 1},\sigma,1}
+ h.c.
\end{equation}
where $W(\bm{r})$ is the hopping amplitude which decays with in-plane distance $r$.
Inserting the inverse Fourier transform, we have
\begin{equation}
H_{\text{inter}} = \frac{1}{N}\sum_{\sigma^\prime \sigma} \sum_{\bm{R}_{\sigma 1} \bm{R}_{\sigma 2}} 
\sum_{\bm{k}^\prime\in\text{BZ}_2}\sum_{\bm{k}\in\text{BZ}_1} e^{-i\bm{k}^\prime \cdot\bm{R}_{\sigma^\prime 2}+i\bm{k}\cdot \bm{R}_{\sigma 1}}
W(\bm{R}_{\sigma^\prime 2}-\bm{R}_{\sigma 1})\ket{\bm{k}^\prime,\sigma^\prime,2} \bra{\bm{k},\sigma,1}
+ h.c.
\end{equation}

On a finite periodic system, the tunneling amplitude $W(\bm{r})$ should also be a periodic function of $\bm{r}$.  
This means we should replace $W(\bm{r})\rightarrow W_{\text{per}}(\bm{r})$ with some periodic function $W_{\text{per}}(\bm{r}+\bm{L})=W_{\text{per}}(\bm{r})$ where $\bm{L}$ is any vector connecting a site with its periodic image.
A periodic tunneling can be constructed as $W_{\text{per}}(\bm{r})=\sum_{\bm{L}} W(\bm{r}+\bm{L})$ where the sum is over the lattice of all $\bm{L}$.
Define the continuum Fourier transform for the original hopping amplitude
\begin{equation}
\begin{split}
\tilde{W}(\bm{q})=\frac{1}{A_{\text{uc}}}\int d\bm{r} e^{-i\bm{q}\cdot\bm{r}} W(\bm{r})\\
W(\bm{r})= A_{\text{uc}}\int \frac{d\bm{q}}{4\pi^2} e^{i\bm{q}\cdot\bm{r}} \tilde{W}(\bm{q})
\end{split}
\end{equation}
where $A_{\text{uc}}$ is the monolayer graphene unit cell area.
This leads to
\begin{equation}
\begin{split}
W_{\text{per}}(\bm{r})
= \sum_{\bm{L}} W(\bm{r}+\bm{L})
= A_{\text{uc}}\sum_{\bm{L}} \int \frac{d\bm{q}}{4\pi^2} e^{i\bm{q}\cdot\bm{L}} e^{i\bm{q}\cdot\bm{r}} \tilde{W}(\bm{q})
 = \frac{1}{N} \sum_{\bm{q}} \tilde{W}(\bm{q}) e^{i\bm{q}\cdot\bm{r}}
\end{split}
\end{equation}
where the sum over $\bm{q}$ in the final line is over the discrete momenta commensurate with the system size $\bm{L}$, and
we have used $\sum_{\bm{L}}e^{i\bm{k}\cdot\bm{L}}=\frac{1}{N A_{\text{uc}}}\sum_{\bm{q}}4\pi^2 \delta^{(2)}(\bm{k}-\bm{q})$, with $\delta^{(2)}$ the two-dimensional Dirac delta function.

The interlayer hopping can then be expressed
\begin{equation}
H_{\text{inter}} = \frac{1}{N^2}\sum_{\bm{q}} \sum_{\sigma^\prime \sigma} \sum_{\bm{R}_{\sigma 1} \bm{R}_{\sigma 2}} 
\sum_{\bm{k}^\prime\in\text{BZ}_2}\sum_{\bm{k}\in\text{BZ}_1} e^{-i(\bm{k}^\prime-\bm{q}) \cdot\bm{R}_{\sigma^\prime 2}+i(\bm{k}-\bm{q})\cdot \bm{R}_{\sigma 1}}
\tilde{W}(\bm{q})\ket{\bm{k}^\prime,\sigma^\prime,2} \bra{\bm{k},\sigma,1}
+ h.c.
\end{equation}
Next, we can use that $\sum_{\bm{R}_{\sigma \ell}} e^{i\bm{k}\cdot \bm{R}_{\sigma \ell}} = N e^{i\bm{k}\cdot\boldsymbol{\tau}_{\sigma \ell}}\sum_{\bm{G}_{\ell}} \delta_{\bm{k},\bm{G}_\ell}$ where $\boldsymbol{\tau}_{\sigma \ell}$ is the offset of the lattice of $\bm{R}_{\sigma \ell}$ from the origin (for graphene, this encodes the offset between the $A$ and $B$ sublattices).
This results in a sum over $\bm{G}_1$ and $\bm{G}_{2}$ and Kronecker deltas $\delta_{\bm{k}^\prime-\bm{q},\bm{G}_2}$ and $\delta_{\bm{k}-\bm{q},\bm{G}_1}$,
\begin{equation}
H_{\text{inter}} = \sum_{\sigma^\prime \sigma} 
\sum_{\bm{q}}
\sum_{\bm{G}_2,\bm{G}_1}
\sum_{\bm{k}^\prime\in\text{BZ}_2}\sum_{\bm{k}\in\text{BZ}_1} 
e^{-i \bm{G}_2 \cdot \boldsymbol{\tau}_{\sigma^\prime 2} + i \bm{G}_1 \cdot \boldsymbol{\tau}_{\sigma 1}}
\delta_{\bm{k}^\prime-\bm{q},\bm{G}_2}\delta_{\bm{k}-\bm{q},\bm{G}_1}
\tilde{W}(\bm{q})\ket{\bm{k}^\prime,\sigma^\prime,2} \bra{\bm{k},\sigma,1}
+ h.c.
\end{equation}
Performing the sum over $\bm{q}$ and relabeling $\bm{G}_{\ell}\rightarrow -\bm{G}_{\ell}$ in the sum finally leads to 
\begin{equation}
H_{\text{inter}} = \sum_{\sigma^\prime \sigma} 
\sum_{\bm{G}_2,\bm{G}_1}
\sum_{\bm{k}^\prime\in\text{BZ}_2}\sum_{\bm{k}\in\text{BZ}_1} 
e^{i \bm{G}_2 \cdot \boldsymbol{\tau}_{\sigma^\prime 2} - i \bm{G}_1 \cdot \boldsymbol{\tau}_{\sigma 1}}
\delta_{\bm{k}+\bm{G}_1,\bm{k}^\prime+\bm{G}_2}
\tilde{W}(\bm{k}+\bm{G_1})\ket{\bm{k}^\prime,\sigma^\prime,2} \bra{\bm{k},\sigma,1}
+ h.c.
\label{eq:Hinter_final}
\end{equation}
Note that for physically realistic tunneling, $\tilde{W}(\bm{q})\rightarrow 0$ as $|\bm{q}|\rightarrow\infty$ so the physics will be dominated tunneling events that occur with the smallest $|\bm{k}+\bm{G}_1|=|\bm{k}^\prime+\bm{G}_2|$.

Our discussion until now has been quite general for two-layer systems.
Let us now specialize to graphene and focus on the low energy physics near the $K$ points and small twist angle $\theta$.

Let us define $\bm{K}=\frac{4\pi}{3a_0}(1,0)$ to be the $K$-point for an untwisted monolayer graphene, and $\bm{b}_1=\frac{4\pi}{\sqrt{3}a_0}(-\frac{\sqrt{3}}{2},\frac{1}{2})$ and $\bm{b}_2=\frac{4\pi}{\sqrt{3}a_0}(-\frac{\sqrt{3}}{2},-\frac{1}{2})$ to be two generators of the reciprocal lattice, i.e. $\bm{G}=n_1\bm{b}_1+n_2\bm{b}_2$ with integers $n_1,n_2$.
They are defined such that
the $C_{3z}$ symmetry related $\bm{K}$ points are
\begin{equation}
\bm{R}(\tfrac{2\pi}{3})\bm{K}=\bm{K}+\bm{b}_1;\;\;\;\;
\bm{R}(\tfrac{4\pi}{3})\bm{K}=\bm{K}+\bm{b}_2
\end{equation}
where $\bm{R}(\phi)$ is the counter-clockwise rotation matrix by angle $\phi$.
We define the layer $\ell$ $K$-point to be $\bm{K}_\ell = \bm{R}(\theta_\ell)\bm{K}$, and reciprocal lattice vectors $(\bm{b}_{\ell 1},\bm{b}_{\ell 2}) = \bm{R}(\theta_\ell)(\bm{b}_1, \bm{b}_2)$, with $\theta_1 = -\theta_2 = \theta/2$.

The low energy physics in graphene occurs near the $\bm{K}$ points.
Let us parameterize momenta near the $\bm{K}_\ell$ point by $\bm{p}_\ell=\bm{k}-\bm{K}_\ell$, where $|\bm{p}_{\ell}|$ is always much smaller than any (non-zero) reciprocal lattice vector.  
Consider the state $\ket{\bm{k},\sigma,1}$ near the $\bm{K}_{1}$ point.
To find all the terms that couple to it, note that for any given $\bm{G}_1$, both $\bm{k}^\prime$ and $\bm{G}_2$ are uniquely fixed by the Kronecker delta $\bm{k}+\bm{G}_1=\bm{k}^\prime+\bm{G}_2$ since $\bm{k}^\prime\in\mathrm{BZ}_2$ lives only in the first BZ.
So, we only need to consider the possible choices of $\bm{G}_1$, for which we can assign a unique tunneling amplitude $\tilde{W}(\bm{k}+\bm{G}_1)$.
There are three choices for which $|\bm{k}+\bm{G}_1|$ is smallest, known as the ``first harmonics'':
\begin{equation}
\begin{split}
(0): \bm{G}_1=0 &\implies |\bm{k}+\bm{G}_1| = |\bm{p}_1+\bm{K}_1|\\
(1): \bm{G}_1=\bm{b}_{1,1}
&\implies |\bm{k}+\bm{G}_1| = |\bm{p}_1+\bm{R}(\tfrac{2\pi}{3})\bm{K}_1|\\
(2): \bm{G}_1=\bm{b}_{1,2}
&\implies |\bm{k}+\bm{G}_1| = |\bm{p}_1+\bm{R}(\tfrac{4\pi}{3})\bm{K}_1|.
\label{eq:firstharmonics}
\end{split}
\end{equation}

The hopping matrix elements in Eq.~\ref{eq:Hinter_final} send $\ket{\bm{p}_{1}+\bm{K}_1,\sigma,1}\rightarrow\ket{\bm{p}_1+\bm{K}_1+\bm{G}_1-\bm{G}_2,\sigma^\prime,2}\equiv\ket{\bm{p}_2+\bm{K}_2,\sigma^\prime,2}$
where $\bm{G}_2$ should be chosen so that $\bm{p}_2$ remains a small momentum.
This amounts to letting $\bm{G}_2=0,\bm{b}_{2,1},$ and $\bm{b}_{2,2}$ for the three tunneling terms listed above.
These three tunneling matrix elements, labeled $n=0,1,2,$ therefore send $\bm{p}_1\rightarrow \bm{p}_2=\bm{p}_1+\bm{K}_1-\bm{K}_2+\bm{G}_1-\bm{G}_2\equiv \bm{p}_1 + \bm{q}_n$ where 
\begin{equation}
\bm{q}_0=\bm{K}_1-\bm{K}_2;\;\;
\bm{q}_1=\bm{K}_1-\bm{K}_2 + \bm{b}_{1,1}-\bm{b}_{2,1};\;\;
\bm{q}_2=\bm{K}_1-\bm{K}_2 + \bm{b}_{1,2}-\bm{b}_{2,2}
\end{equation}
characterize the three first harmonic tunneling terms.
Notice that $\bm{q}_n=\bm{R}(\tfrac{2\pi n}{3})\bm{q}_0$ are related to each other by rotations.
From Eq.~\ref{eq:firstharmonics}, these three terms occur with the tunneling amplitudes $\tilde{W}(\bm{p}_1+\bm{R}(\tfrac{2\pi n}{3})\bm{K}_1)$.

The Bistritzer-MacDonald continuum model follows from keeping only these three first harmonic terms and approximating the tunneling amplitude by its value at $\bm{K}$: $\tilde{W}(\bm{p}_1+\bm{R}(\tfrac{2\pi n}{3})\bm{K}_1) \approx \tilde{W}(\bm{K})$.
Since the tunneling amplitude does not depend on $\bm{p}_1$, it is ``momentum independent''.
The momentum dependent tunneling (MDT) term used in the main text incorporates the next leading order correction in this approximation.
We now illustrate how this is taken into account.

Let $\bar{\bm{K}}=(\bm{K}_1+\bm{K}_2)/2$ be an average of the two $\bm{K}$-points (for $N_L$ layers, one would choose the average of all $N_L$).
For each of the three tunneling events, we Taylor expand $\tilde{W}(\bm{q})$ near $\bm{q}=\bm{R}(\tfrac{2\pi n}{3})\bar{\bm{K}}$.
The tunneling amplitude then takes the form
\begin{equation}
\tilde{W}(\bm{p}_1+\bm{R}(\tfrac{2\pi n}{3})\bm{K}_1)\approx\tilde{W}(\bm{R}(\tfrac{2\pi n}{3})\bar{\bm{K}})+ (\bm{p}_1+\bm{R}(\tfrac{2\pi n}{3})(\bm{K}_1-\bar{\bm{K}}))\cdot \bm{\nabla}\tilde{W}(\bm{R}(\tfrac{2\pi n}{3})\bar{\bm{K}})
\label{eq:ttilde}
\end{equation}
where $\bm{\nabla}\tilde{W}$ is the gradient of $\tilde{W}$.

In twisted bilayer graphene and h-HTG, $\bar{\bm{K}}$ points along the $x$-direction and $\bm{q}_0$ points along the $y$-direction.
If we take $\tilde{W}(\bm{k})$ to be rotation symmetric, then it only depends on $|\bm{k}|$, and therefore $\bm{\nabla}\tilde{W}(\bar{\bm{K}})$ points in the $x$ direction.
Since the three tunneling terms are related by $2\pi/3$ rotations, we see that the tunneling term with $\bm{q}_n$ only depends on the magnitude of $\bm{p}_1$ perpendicular to $\bm{q}_n$.
Defining $\hat{\bm{q}}_{n,\perp}=(\cos\tfrac{2\pi n}{3},\sin\tfrac{2\pi n}{3})$,
Eq~\ref{eq:ttilde} simplifies to
\begin{equation}
\tilde{W}(\bm{p}_1+\bm{R}(\tfrac{2\pi n}{3})\bm{K}_1)\approx w_0(1+\lambda_{\mathrm{MDT}}  \hat{\bm{q}}_{n,\perp}\cdot \bm{p}_1)
\end{equation}
as quoted in the main text.
In terms of $\tilde{W}(\bm{q})$, we have $w_0=\tilde{W}(\bar{\bm{K}})$ which is the usual (momentum-independent) interlayer tunneling strength in the Bistritzer-MacDonald continuum model, and
$\lambda_{\mathrm{MDT}}= (\bm{\nabla}\tilde{W}(\bar{\bm{K}}))_x/w_0$
accounts for the leading order correction.

\begin{figure}[h!]
    \centering\includegraphics[width=0.4\linewidth]{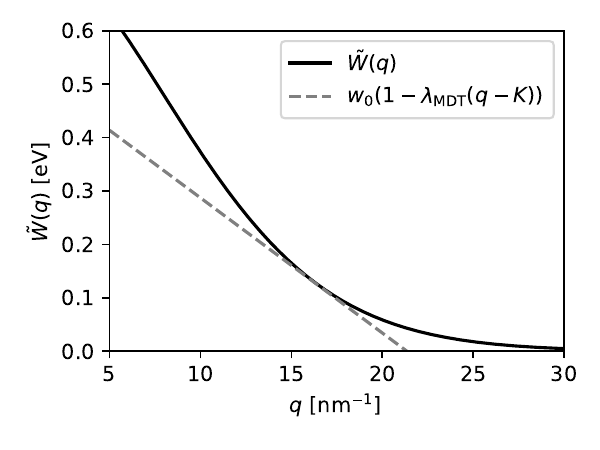}
    \caption{Plot of $\tilde{W}(q)$, the Fourier transform on the interlayer tunneling amplitude.
    Dashed line shows the linear approximation which is accurate near the $K$ point of graphene.
    }
    \label{appfig_Tq}
\end{figure}

To estimate the magnitude of $\lambda_{\text{MDT}}$, we plot in Fig~\ref{appfig_Tq} the tunneling amplitude for the Slater-Koster~\cite{slater1954simplified} form for $W(\bm{r})$ with the parameters used in Ref~\cite{moon2013optical,koshino2015interlayer}.  
We also plot the linear approximation, using $w_0=w_{AB}=110$meV and $\lambda_{\mathrm{MDT}}=-2.3$\AA{} as used in the main text for the AB interlayer tunneling. 
Indeed, this approximation captures the behavior of $\tilde{W}(q)$ near the $K$-point at $q\approx 17$nm$^{-1}$. 

Due to lattice relaxation effects at the moir\'e scale, the effective inter-sublattice and intra-sublattice tunneling strengths differ. In this work, we incorporate this by scaling the inter-sublattice tunneling by a factor $w_{AA}/w_{AB}$, where we take $w_{AA}=75\,\text{meV}$.

We note that the precise value of $\lambda_{\mathrm{MDT}}$ does not qualitatively affect the interacting results in this work. For example, in Fig.~\ref{figapp:HFsubpol_additional}, we compare HF calculations using $\lambda_{\mathrm{MDT}}=-2.0$\AA{} and $-2.3$\AA{}. 

\section{Single-particle band structure with and without inter-Chern hybridization}\label{secapp:interChern}

\begin{figure*}[t!]
    \centering
    \includegraphics[width=1\linewidth]{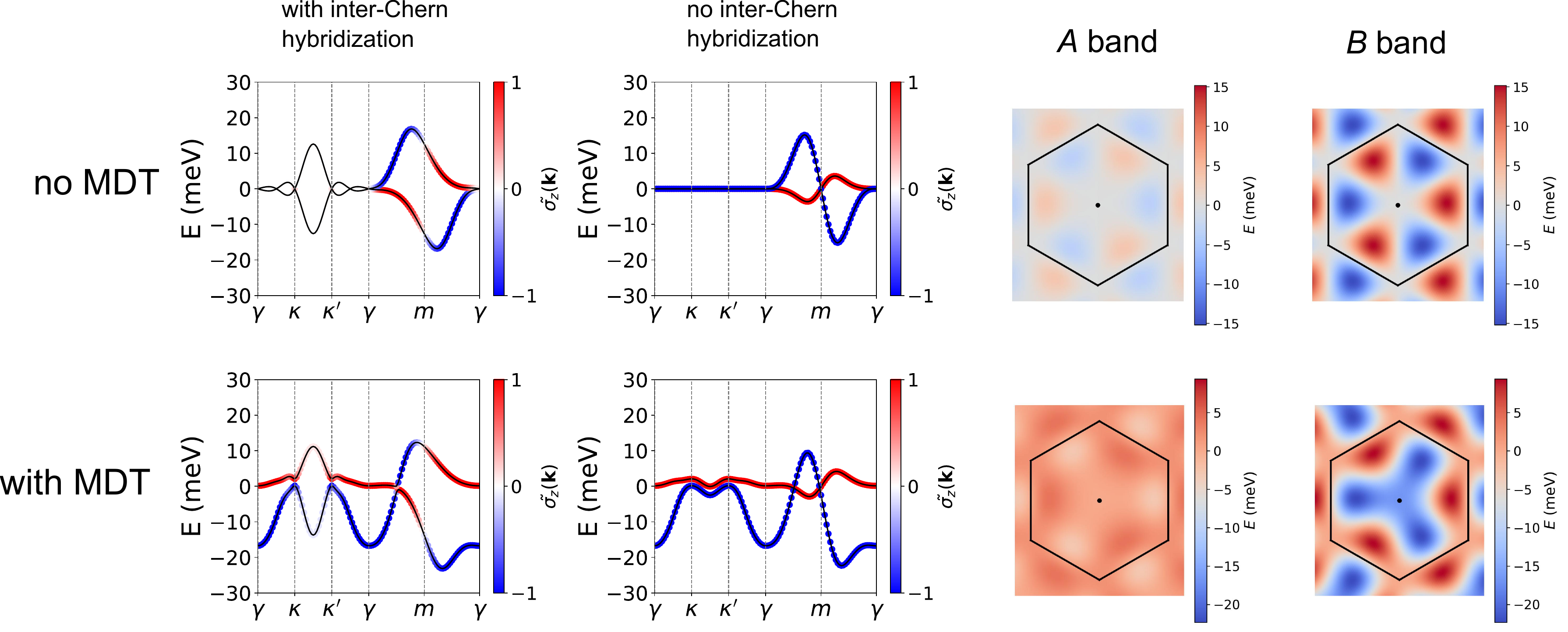}
    \caption{\textbf{Single-particle band structure with and without inter-Chern hybridization at $U=0\,$meV.} Top (bottom) row shows results without (with) MDT. The first and second columns show the dispersion of the central bands with and without inter-Chern hybridization. The color indicates the Chern-sublattice polarization. The third and fourth columns show the $A$ and $B$ band dispersions in the mBZ (black hexagon) without the inter-Chern hybridization.  Without MDT, $E_A(\bm{k})$ has a mean energy of $0.0\,$meV and is bounded by $-3.6\,$meV and $3.6\,$meV, while $E_B(\bm{k})$ has a mean energy of $0,0\,$meV and is bounded by $-15.2\,$meV and $15.2\,$meV. With MDT, $E_A(\bm{k})$ has a mean energy of $1.0\,$meV and is bounded by $-2.9\,$meV and $4.2\,$meV, while $E_B(\bm{k})$ has a mean energy of $-6.4\,$meV and is bounded by $-22.4\,$meV and $9.4\,$meV. Twist angle is $\theta=1.8^\circ$, and interlayer potential is $U=0$\,meV.}
    \label{figapp:interChern}
\end{figure*}

In Fig.~\ref{figapp:interChern}, we show additional details of the dispersion of the central bands at $\theta=1.8^\circ$ and vanishing interlayer potential $U=0\,$meV, comparing the cases with and without MDT. In particular, we also consider switching off the inter-Chern hybridization, i.e.~we rotate the single-particle dispersion to the Chern-sublattice basis, and neglect the off-diagonal terms in Chern-sublattice space. This procedure provides a concrete definition of the dispersions $E_{\tilde{\sigma}_z}(\bm{k})$ of the $A$ and $B$ bands. It can be seen that the $B$ band is significantly more dispersive than the $A$ band regardless of MDT. With MDT though, the $B$ band is generally shifted down compared to the $A$ band, especially near center of the mBZ. The quantum geometric properties of the Chern-sublattice bands are robust to MDT owing to the large remote gaps. 

\begin{figure*}[t!]
    \centering
    \includegraphics[width=1\linewidth]{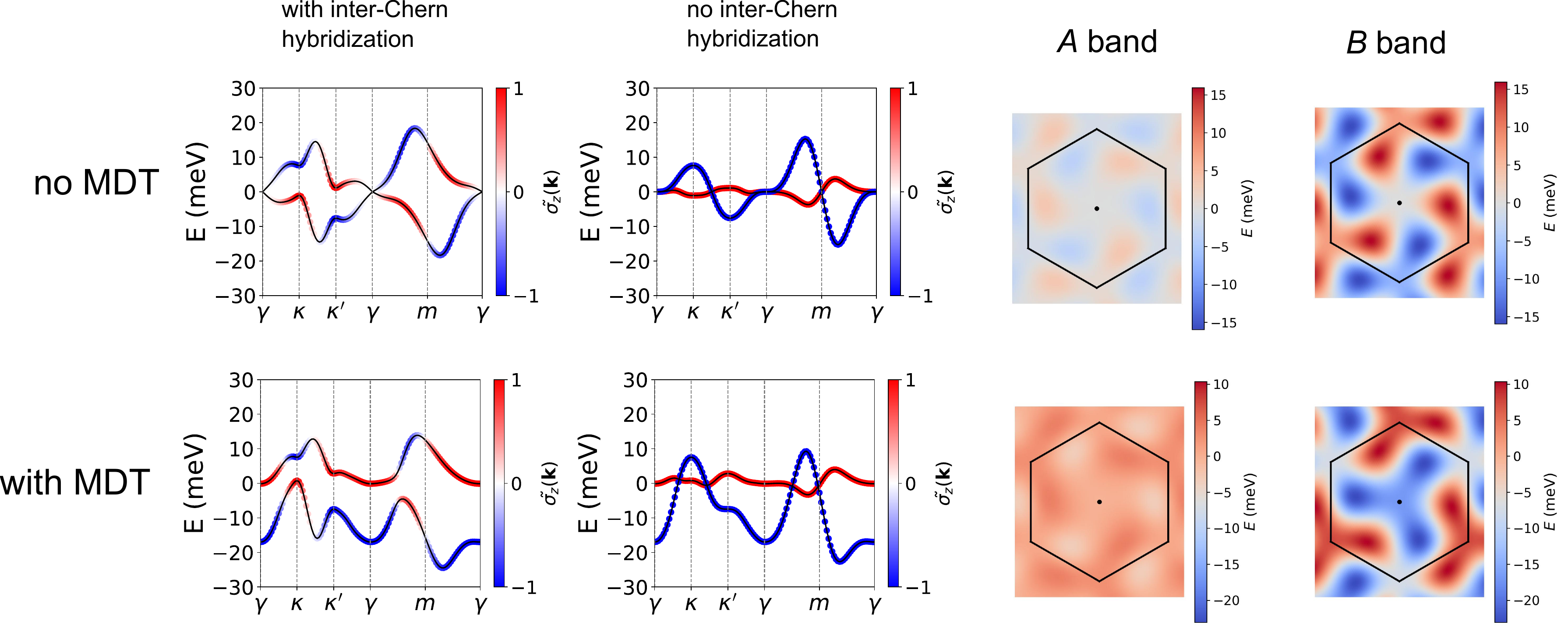}
    \caption{\textbf{Single-particle band structure with and without inter-Chern hybridization at $U=20\,$meV.} Top (bottom) row shows results without (with) MDT. The first and second columns show the dispersion of the central bands with and without inter-Chern hybridization. The color indicates the Chern-sublattice polarization. The third and fourth columns show the $A$ and $B$ band dispersions in the mBZ (black hexagon) without the inter-Chern hybridization.  Without MDT, $E_A(\bm{k})$ has a mean energy of $0.0\,$meV and is bounded by $-3.7\,$meV and $3.7\,$meV, while $E_B(\bm{k})$ has a mean energy of $0,0\,$meV and is bounded by $-16.0\,$meV and $16.0\,$meV. With MDT, $E_A(\bm{k})$ has a mean energy of $0.7\,$meV and is bounded by $-3.3\,$meV and $4.0\,$meV, while $E_B(\bm{k})$ has a mean energy of $-6.6\,$meV and is bounded by $-23.1\,$meV and $10.4\,$meV. Twist angle is $\theta=1.8^\circ$, and interlayer potential is $U=20$\,meV.}
    \label{figapp:interChern_U0.020}
\end{figure*}

 Fig.~\ref{figapp:interChern_U0.020} shows analogous results but with interlayer potential $U=20\,$meV. Note that the interlayer potential $U$ acts as a sublattice diagonal onsite potential with strength $U,0,-U$ on the top, middle, and bottom layers respectively. Regardless of MDT, the main effect of the displacement field is to raise (lower) the energy of the $B$ band at the mBZ corner $\kappa$ ($\kappa'$).

In all plots of single-particle dispersions in this work, we show the band structure relative to the charge neutrality point. For all interacting calculations in this work, we use the full single-particle dispersion that includes the hybridization between the $A$ and $B$ bands.

\clearpage

\section{Additional Hartree-Fock calculations}

\subsection{Phase diagrams for other integer fillings}

\begin{figure*}[t!]
    \centering
    \includegraphics[width=1\linewidth]{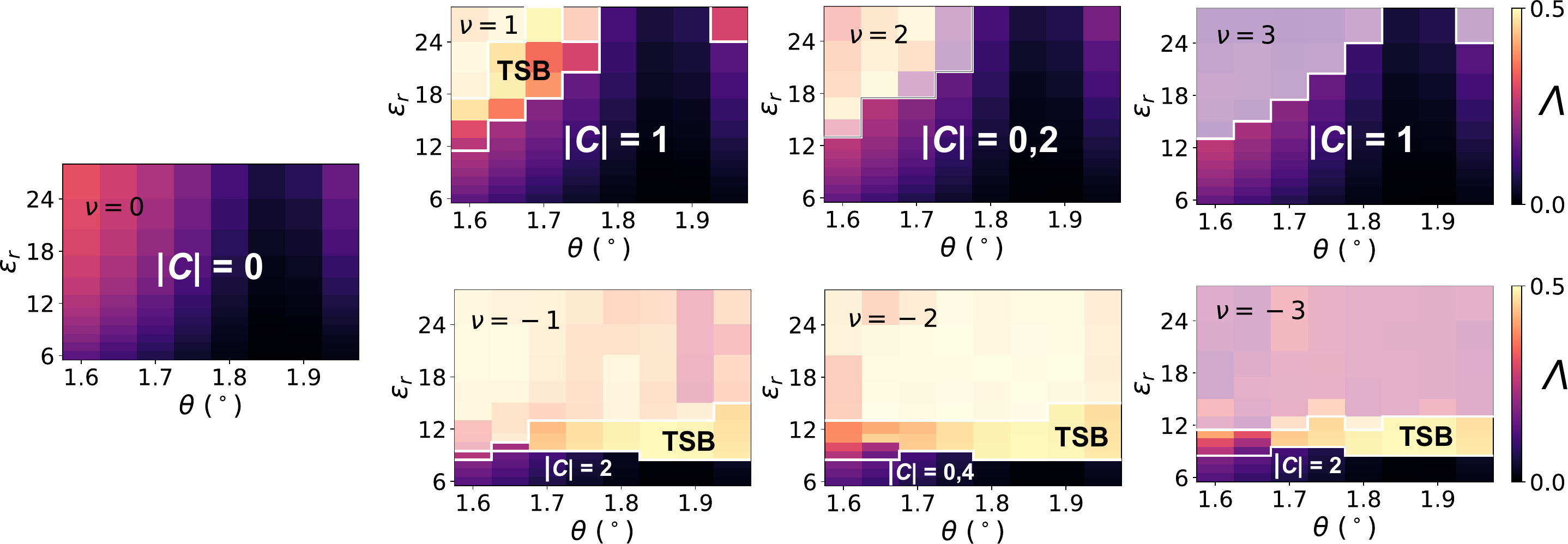}
    \caption{\textbf{Hartree-Fock phase diagram with MDT at all integer fillings.} HF phase diagram as a function of twist angle $\theta$ and relative permittivity $\epsilon_r$. White shaded regions indicate gapless states. $\Lambda\equiv \text{max}_{\tau,s,\sigma}\Big[\text{min}\big(\langle n_{\tau,s,\sigma}\rangle,1-\langle n_{\tau,s,\sigma}\rangle\big)\Big]$ is a measure of how far the HF solution is from an idealized strong-coupling state. TSB indicates various states that break moir\'e translation symmetry. System size is $12\times 12$, and we allow for spontaneous doubling or tripling of the moir\'e unit cell along both directions. Interlayer potential is $U=0$.}
    \label{figapp:HF_phase_diagram_all_int}
\end{figure*}

In Fig.~\ref{figapp:HF_phase_diagram_all_int}, we show HF phase diagrams for all non-integer fillings in the presence of MDT. The particle-hole asymmetry is apparent for all non-integer fillings.  For positive integer fillings, the strong-coupling insulator occupies the majority of the phase diagram, while for negative integer fillings, it only survives for strong interactions. The Chern numbers can be understood from the principle that starting from band insulator at $\nu=-4$ (with $C=0$), the system first occupies $B$ bands with $|C|=2$, then $A$ bands with $|C|=1$.  For $|\nu|=2$, our HF calculations find that strong-coupling states that differ by the valley polarization are exactly degenerate, which arises from a `flavor permutation symmetry' that holds at mean-field level~\cite{kwan2024strong}. For $\nu=0$, no symmetries are broken, and the system can sustain an indirect HF gap for rather weak interactions. 

\subsection{Quantum geometry of the $|C|=1$ band at $\nu=+3$}

\begin{figure*}[h!]
    \centering\includegraphics[width=0.48\linewidth]{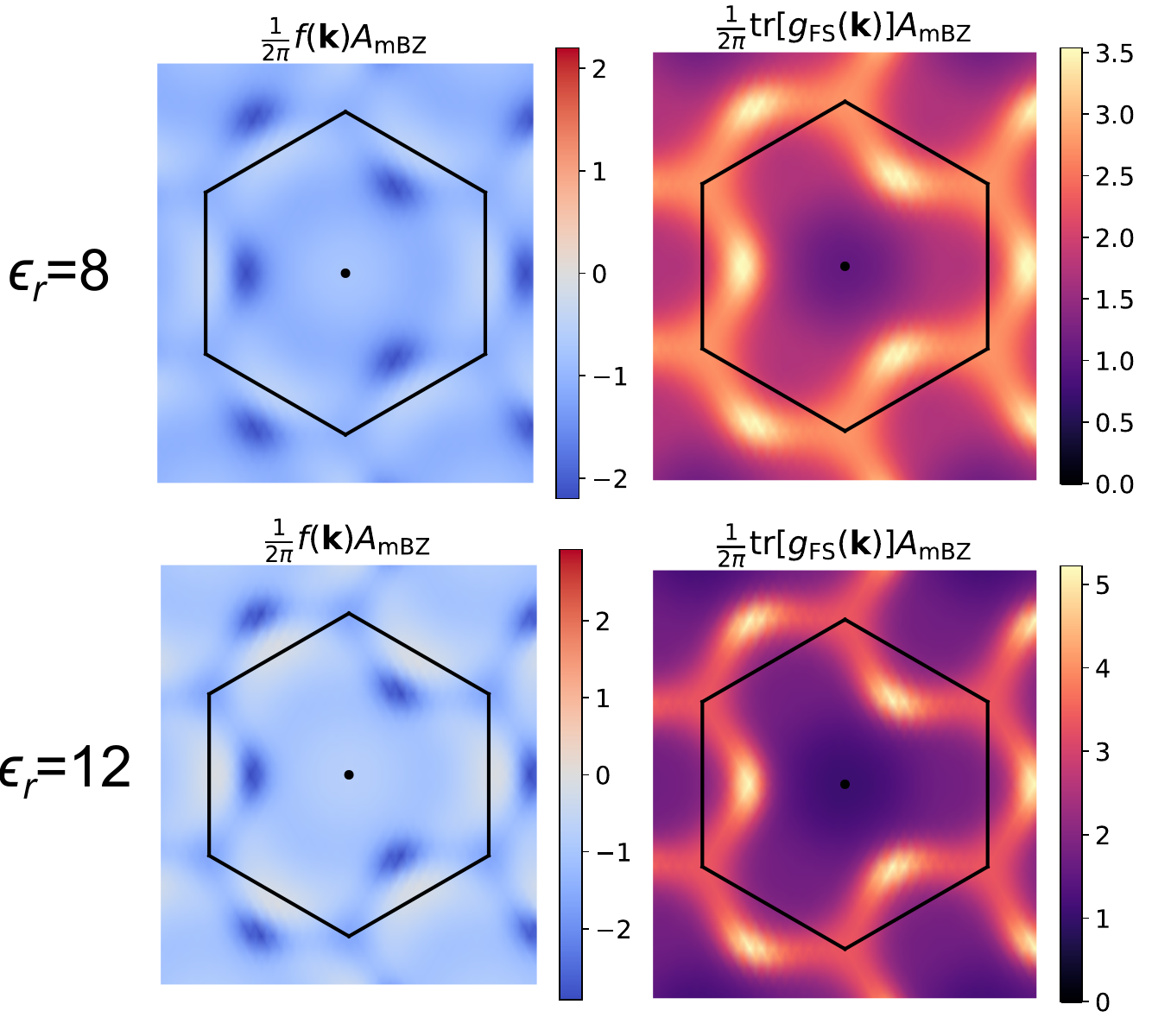}
    \caption{Quantum geometry of the unoccupied HF $|C|=1$ band at $\nu=+3$ for $\epsilon_r=8,12$. $f(\bm{k})$ is the Berry curvature, and $g_\text{FS}(\bm{k})$ is the Fubini-Study metric. The trace condition violation $\frac{A_{\text{mBZ}}}{2\pi}\int d\bm{k}\,\text{tr}\left([g_{\text{FS}}(\bm{k})]-|f(\bm{k})|\right)$, where $A_{\text{mBZ}}$ is the momentum area of the mBZ, is 1.05 and 1.30 for $\epsilon_r=8,15$ respectively. System size is $24\times 24$, $\theta=1.80^\circ$, and $U=0$.}
    \label{figapp:HF_QG}
\end{figure*}

In Fig.~\ref{figapp:HF_QG}, we compute the Berry curvature $f(\bm{k})$ and the trace of the Fubini-Study metric $g_\text{FS}(\bm{k})$~\cite{parameswaranFractionalQuantumHall2013} for the unoccupied $|C|=1$ HF band of the strong-coupling insulator at $\nu=+3$. The curvature is relatively homogeneous throughout the mBZ, and the violation of the trace condition $\frac{A_{\text{mBZ}}}{2\pi}\int d\bm{k}\,\text{tr}\left([g_{\text{FS}}(\bm{k})]-|f(\bm{k})|\right)$, where $A_{\text{mBZ}}$, is the area of the mBZ, is fairly small. The quantum geometry becomes smoother as the interaction strength is increased. 

\subsection{Additional HF data for $3\leq \nu\leq 4$ Chern-sublattice polarization}
\begin{figure*}[t!]
    \centering
    \includegraphics[width=0.5\linewidth]{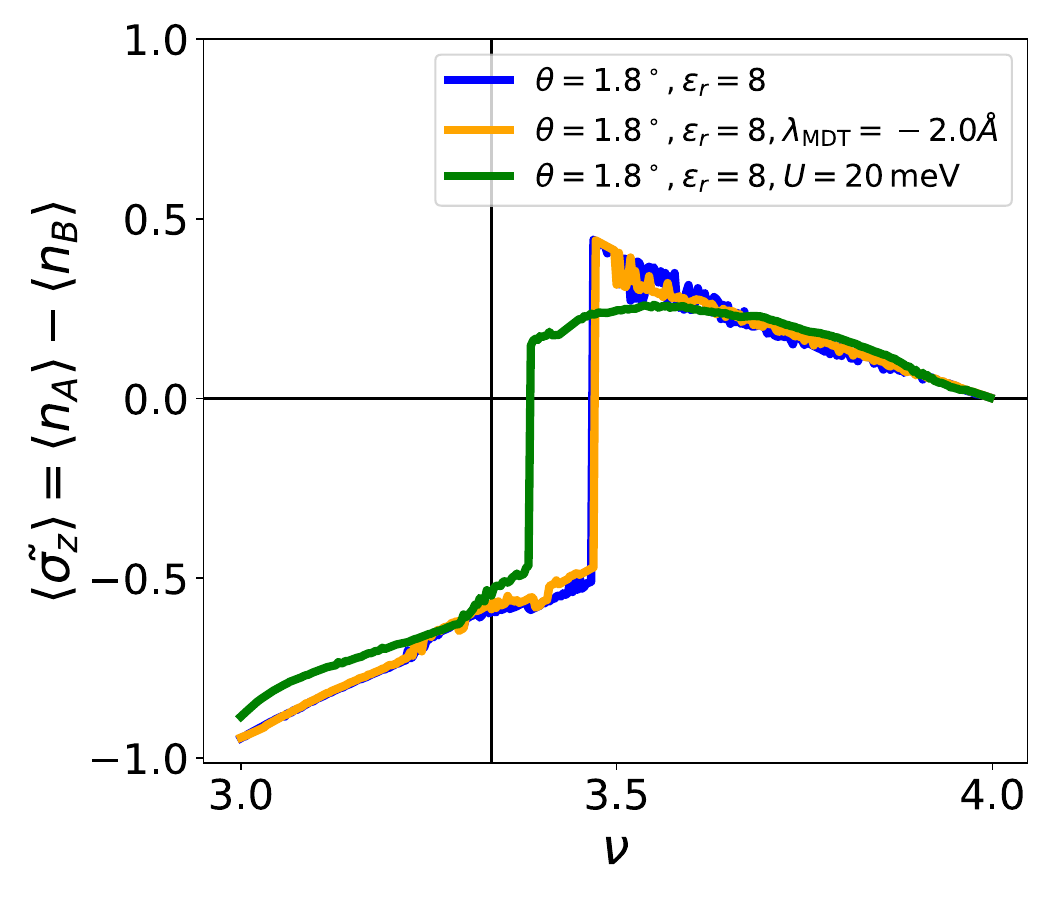}
    \caption{\textbf{Additional data for 
    Chern-sublattice polarization in HF.} System size is $18\times 18$}
    \label{figapp:HFsubpol_additional}
\end{figure*}

In Fig.~\ref{figapp:HFsubpol_additional}, we show the Chern-sublattice polarization $\langle \tilde{\sigma}_z\rangle $ obtained with HF calculations in the filling range $3\leq \nu\leq 4$ for additional sets of parameters. The data for $\lambda_{\text{MDT}}=-2.0\,$\r{A} is remarkably similar to that for $\lambda_{\text{MDT}}=-2.3\,$\r{A}, which is the default value used in this paper. Application of a moderate interlayer potential $U=20\,$meV leads to a reduction of $\nu_c$.

\subsection{Properties at all fillings}\label{subsecapp:HF_noninteger}

\begin{figure*}[t!]
    \centering
    \includegraphics[width=1\linewidth]{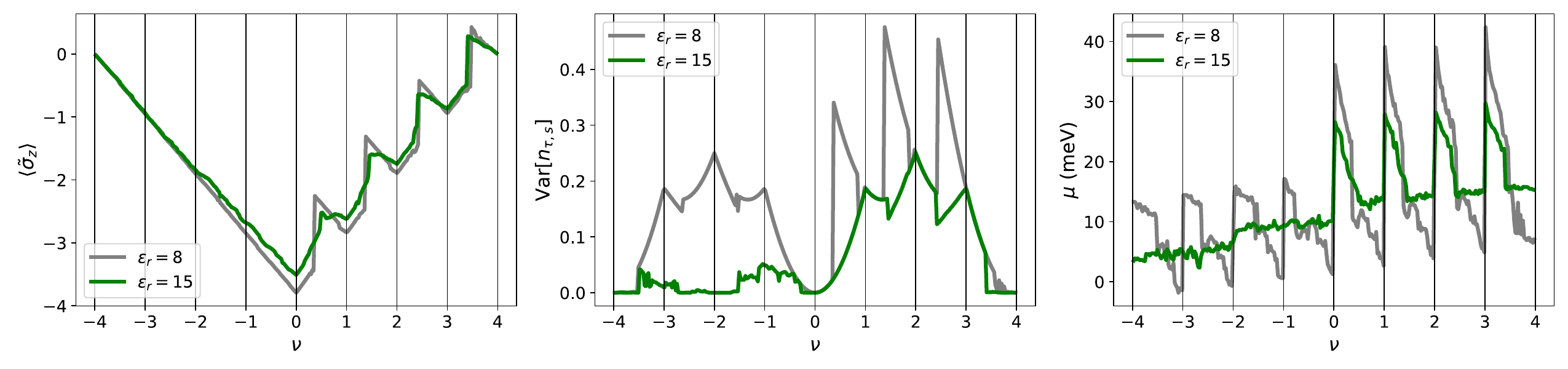}
    \caption{\textbf{Properties of the Hartree-Fock solution with MDT as a function of $\nu$.} We show results for $\epsilon_r=8,15$. Left: Chern sublattice polarization $\langle \tilde{\sigma}_z\rangle$. Middle: Variance in the flavor occupations $\text{Var}[n_{\tau,s}]$. Right: Chemical potential. System size is $18\times 18$, $\theta=1.80^\circ$, $U=0\,$meV, and translation symmetry and valley $U_v(1)$ is enforced.}
    \label{figapp:allnu}
\end{figure*}

In Fig.~\ref{figapp:allnu}, we present various properties of the HF solution throughout the narrow-band filling regime $-4<\nu<+4$. The sublattice polarization $\langle \tilde{\sigma}_z\rangle$ monotonically decreases from 0 for $-4<\nu<0$, as the system populates the dispersive $B$ bands with overall lower kinetic energy than the $A$ bands. For $\nu>0$, we observe a series of sublattice transitions that repeats between every integer filling. $\mathrm{Var}[n_{\tau,s}]$ is the variance of the flavor occupations (i.e.~the number of particles with that flavor per moir\'e unit cell) across the two valleys $\tau$ and two spins $s$, and gives a measure of the degree of flavor polarization. For $\nu<0$, the $\epsilon_r=15$ calculation shows only weak flavor polarization. For $\nu>0$, the data for $\epsilon_r=8$ and $\epsilon_r=15$ show qualitatively different behavior, which is due to the fact that the former is in the `flavor-imbalanced' regime with stronger interactions, while the latter is in the `flavor-balanced' regime~\cite{kwan2024strong}. However, both interaction strengths yield a fully flavor-polarized state between $\nu=+3$ and the critical filling $\nu_c$ at which the sublattice transition in the interval $3<\nu<4$ occurs. For $\epsilon_r=8$, the full flavor polarization persists for a finite window beyond $\nu_c$, implying that the position of the sublattice transition here is not controlled by the flavor physics. $\mu$ tracks the chemical potential evolution as a function of $\nu$. There are sharp negative features in the inverse compressibility $\sim \frac{d\mu}{d\nu}$, especially at the sublattice transitions. Note that the overall increase in $\mu$ from $\nu=-4$ to $\nu=+4$ is only around $\sim 10\,\text{meV}$, which is an indication that Hartree renormalization effects in h-HTG are not large.

\subsection{HF phase diagram at $\nu=+3$ with interlayer potential}

\begin{figure*}[h!]
    \centering\includegraphics[width=0.48\linewidth]{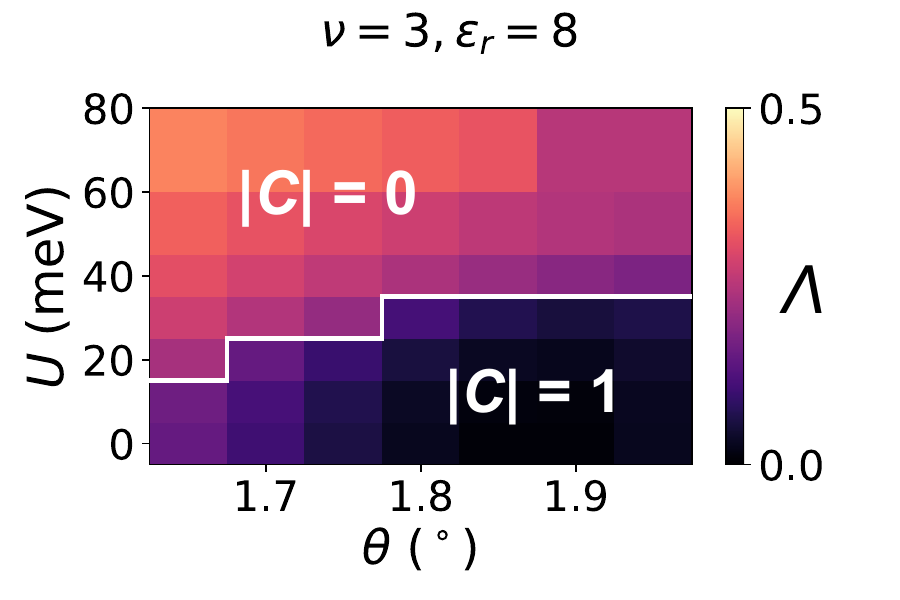}
    \caption{\textbf{Hartree-Fock phase diagram at $\nu=+3$ with MDT and interlayer potential $U$.}  $\Lambda\equiv \text{max}_{\tau,s,\tilde{\sigma}}\Big[\text{min}\big(\langle n_{\tau,s,\tilde{\sigma}}\rangle,1-\langle n_{\tau,s,\tilde{\sigma}}\rangle\big)\Big]$ is a measure of how far the HF solution is from an idealized strong-coupling state. System size is $12\times 12$, we fix $\epsilon_r=8$, and we enforce moir\'e translation invariance.}
    \label{figapp:HF_phase_diagram_U_nu3}
\end{figure*}

In Fig.~\ref{figapp:HF_phase_diagram_U_nu3}, we show the HF phase diagram with $\nu=+3$ in the presence of MDT and an interlayer potential $U$ for $\epsilon_r=8$. For $U\lesssim 30\,\text{meV}$, we find a strong-coupling $|C|=1$ insulator, which undergoes a displacement-field-tuned transition to a $|C|=0$ insulator that does not fit the strong-coupling framework~\cite{kwan2024strong}. This $|C|=0$ state can be understood as arising from momentum-local sublattice inversions. Note that at even higher $U$, we expect to find various intervalley-coherent and translation symmetry-broken phases, but we enforce valley $U(1)_v$ and moir\'e translation symmetry here for simplicity. 

\clearpage

\section{Additional exact diagonalization calculations}\label{secapp:addtional_ED}
\subsection{Calculations for $\theta=1.8^{\circ}$ with displacement field}

\begin{figure*}[h!]
    \centering\includegraphics[width=\linewidth]{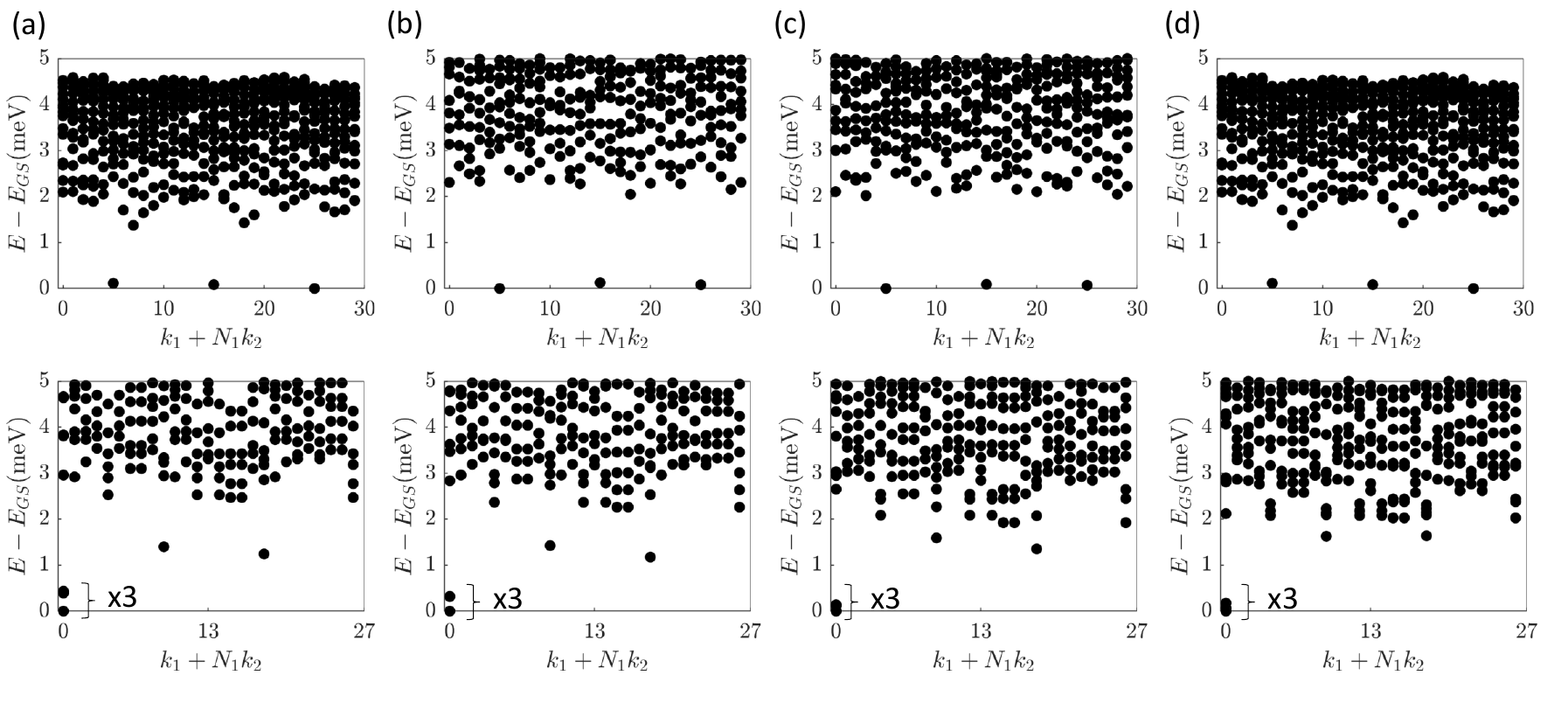}
    \caption{1-band ED calculations at $\nu=3+\frac{1}{3}$ at finite interlayer potential $U$. We fix $\theta=1.8^{\circ}$ and $\epsilon_r=8$ for all plots. Top row is for the $5\times 6$ site cluster, and bottom row is for the tilted 27 site cluster. Interlayer potential $U$ is (a) $5$\,meV, (b) $10$\,meV, (c) $20$\,meV, and (d) $30$\,meV.}
    \label{appfig_withD_t1.8_oneED}
\end{figure*}

\begin{figure*}[h!]
    \centering\includegraphics[width=0.48\linewidth]{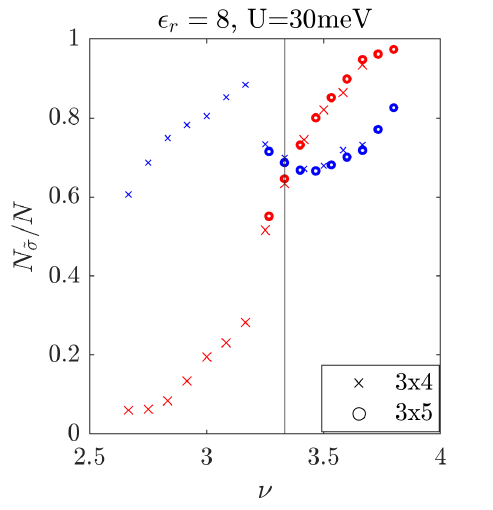}
    \caption{ 2-band ED calculation of the Chern-sublattice polarization as a function of filling. The legend indicates the lattice geometry of the cluster. $N_{\tilde{\sigma}}$ is the expectation value of the number of particles in the Chern-sublattice band $\tilde{\sigma}$. The vertical line corresponds to $\nu=3+1/3$. We fix $\theta=1.8^{\circ}$ and $U=30\,$meV. Red corresponds to the $A$ sublattice, and blue corresponds to the $B$ sublattice. }
    \label{appfig_withD_t1.8_twoED}
\end{figure*}

Fig.~\ref{appfig_withD_t1.8_oneED} and \ref{appfig_withD_t1.8_twoED} show additional $\nu=3+\frac{1}{3}$ ED data performed at $\theta=1.8^{\circ}$ with finite interlayer potential $U$ and $\epsilon_r=8$. For a wide range of displacement fields $|U|<30$ meV, 1-band ED shows good evidence of an FCI. However, Fig.~\ref{appfig_withD_t1.8_twoED} demonstrates that a sufficiently large displacement field can undermine the sublattice-Chern polarization, which makes restriction to a single active HF band less justified.

\subsection{Calculations for $\theta=1.8^{\circ}$ without displacement field}

\begin{figure*}[h!]
    \centering\includegraphics[width=\linewidth]{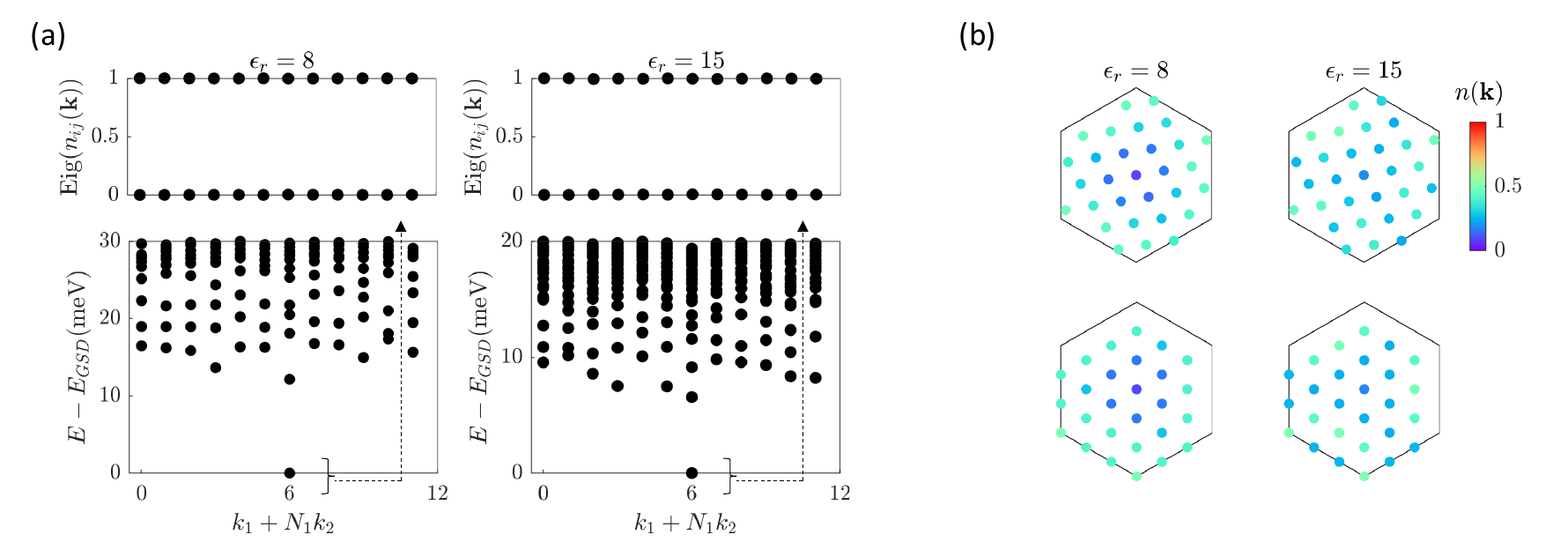}
    \caption{ (a) 2-band ED calculations at $\nu=3$ with $\theta=1.8^{\circ}$ and $U=0$. Bottom panels show the momentum-resolved many-body spectrum. Top panels show the eigenvalues of the occupation matrix $n_{ij}(\bm{k})$ of the ground state. (b) Occupation number $n(\bm{k})$ of the many-body ground state obtained from 1-band ED performed at $\nu=3+1/3$ with $\theta=1.8^{\circ}$ and $U=0$. We average over the three quasi-degenerate ground states. The top row is for the 30-site cluster, while the bottom row is for the tilted 27-site cluster.}
    \label{appfig_noD_t1.8_oneandtwo}
\end{figure*}

Fig.~\ref{appfig_noD_t1.8_oneandtwo}a shows the many-body spectrum of 2-band ED performed on the $3\times4$ cluster at $\nu=3$. There is a single non-degenerate ground state, separated by a neutral gap of around 10\,meV and 5\,meV for $\epsilon_r=8$ and $\epsilon_r=15$ respectively. The top row of Fig.~\ref{appfig_noD_t1.8_oneandtwo}a shows the eigenvalues of the occupation matrix $n_{ij}(\bm{k})$ of this ground state ($i,j$ index the two bands). The two branches of the eigenvalues are close to either 0 or 1, demonstrating that the ground state at $\nu=3$ is effectively a single Slater determinant state that is polarized in the $A$ band.

Fig.~\ref{appfig_noD_t1.8_oneandtwo}b plots the occupation number $n(\bm{k})$ of the ground state obtained from 1-band ED at $\nu=3+1/3$ . There is still considerable fluctuation in $n(\bm{k})$ among different $\bm{k}$. This suggests that the competition from the kinetic energy is still important, and may be an important source of finite-size effects observed in our ED calculations.

\subsection{Calculations for $\theta=1.85^{\circ}$}
\begin{figure*}[h!]
    \centering\includegraphics[width=\linewidth]{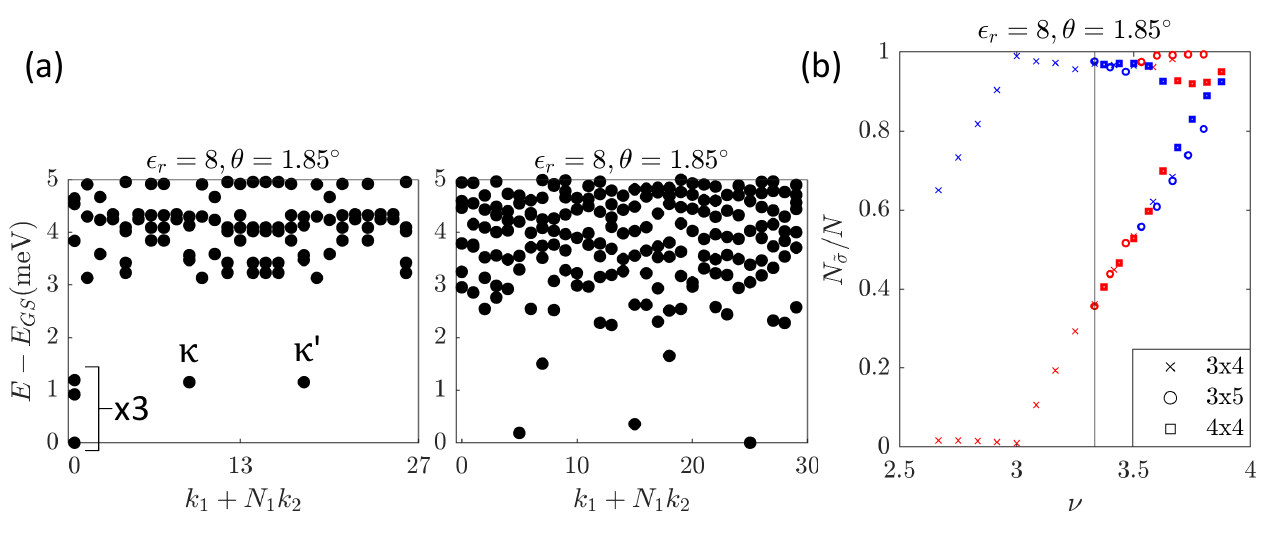}
    \caption{ \textbf{(a)} 1-band ED calculations on the tilted 27-site (left) and the $5\times6$ (right) cluster at $\nu=3+1/3$.  \textbf{(b)} 2-band ED calculation of the Chern-sublattice polarization as a function of filling. Different symbols (see legend) refer to various lattice geometries with corresponding system sizes $N=12,15,16$. $N_{\tilde{\sigma}}$ is the expectation value of the particle occupation in the Chern-sublattice basis. Red (blue) corresponds to the $A$ ($B$) basis. The vertical line indicates $\nu=3+1/3$. All calculations are performed at $\theta=1.85^\circ$, $\epsilon_r=8$, and with MDT.}
    \label{appfig_t1.85}
\end{figure*}

Fig.~\ref{appfig_t1.85} shows additional ED data at twist angle $\theta=1.85^{\circ}$.  1-band ED performed on the $5\times6$ cluster shows clear signs of an FCI, as indicated by the 3-fold quasi-degenerate ground states that evolve into each other and remain gapped from higher states under flux threading. However, calculations performed on the tilted 27-site lattice~\cite{PhysRevB.103.125406} show that there are competing excitations at $\kappa$ and $\kappa'$. This should be compared with the result at $\theta=1.8^{\circ}$, where there are three quasi-degenerate ground states at $\gamma$ and the possibility of charge density wave is excluded.

\subsection{Multi-band calculations of FCIs}\label{secapp:2band_bandmixing}

In this subsection, we discuss multi-band ED calculations of FCIs in the fully flavor polarized sector at $\nu=3+\frac{1}{3}$. Owing to the $U(2)_K\times U(2)_{\bar{K}}$ symmetry of the Hamiltonian, we can focus our attention on the two bands within the single partially-occupied flavor. The main limitation of ED stems from the exponential growth of the many-body Hilbert space. While performing a 1-band calculation within the $\nu=+3$ HF conduction band significantly reduces the computational cost, this restriction ignores band mixing effects with the other band~\cite{yu2024versus,xu2024maximally,abouelkomsan2024mixing,yu2024moirefractionalcherninsulators,kwan2024abelianfractionaltopologicalinsulators}, which are potentially important especially since the two non-interacting bands are not isolated. 

One solution is to simply perform full 2-band calculations, but this severely limits the system sizes $N$ that can be accessed. For example, the Hilbert space dimensions per momentum sector for $N=15,18,21$ are $2\times10^6,7\times 10^7,3\times 10^9$. While $N=18$ is in principle possible, it is out of reach of our current computational resources. 

Another solution is to introduce a `bandmax' constraint~\cite{rezayi2011breaking,yu2024moirefractionalcherninsulators,kwan2024abelianfractionaltopologicalinsulators} that systematically interpolates between 1-band and 2-band calculations, and allows for larger $N$. We consider working in the $\nu=+3$ HF basis of the 2-band space. At $\nu=+3$, HF calculations yield a conduction band (which we refer to as $A$ since it has high polarization on the $A$ sublattice-Chern band) separated by a HF gap to the valence band (which we refer to as $B$). The HF and 2-band ED calculations show that the ground state at $\nu=3+1/3$ continues to have near-complete occupancy of the $B$ band. Hence for multi-band ED calculations at $\nu=3+1/3$, we can constrain the many-body Hilbert space to include Fock states that include at most $N_\text{max}$ holes in the $B$ band. $N_\text{max}=0$ corresponds to the 1-band calculation, while $N_\text{max}=N$ corresponds to the full 2-band calculation. Based on the HF and 2-band ED results, we anticipate a rapid convergence with $N_\text{max}$, so that we can effectively understand the low-energy 2-band physics at relatively small $N_\text{max}$. For $N=18$, an $N_\text{max}=4$ calculation at $\nu=3+\frac{1}{3}$ has a Hilbert space dimension of $1\times 10^7$ per momentum sector, which is within reach of our computational resources. 

\begin{figure*}[h!]
    \centering\includegraphics[width=\linewidth]{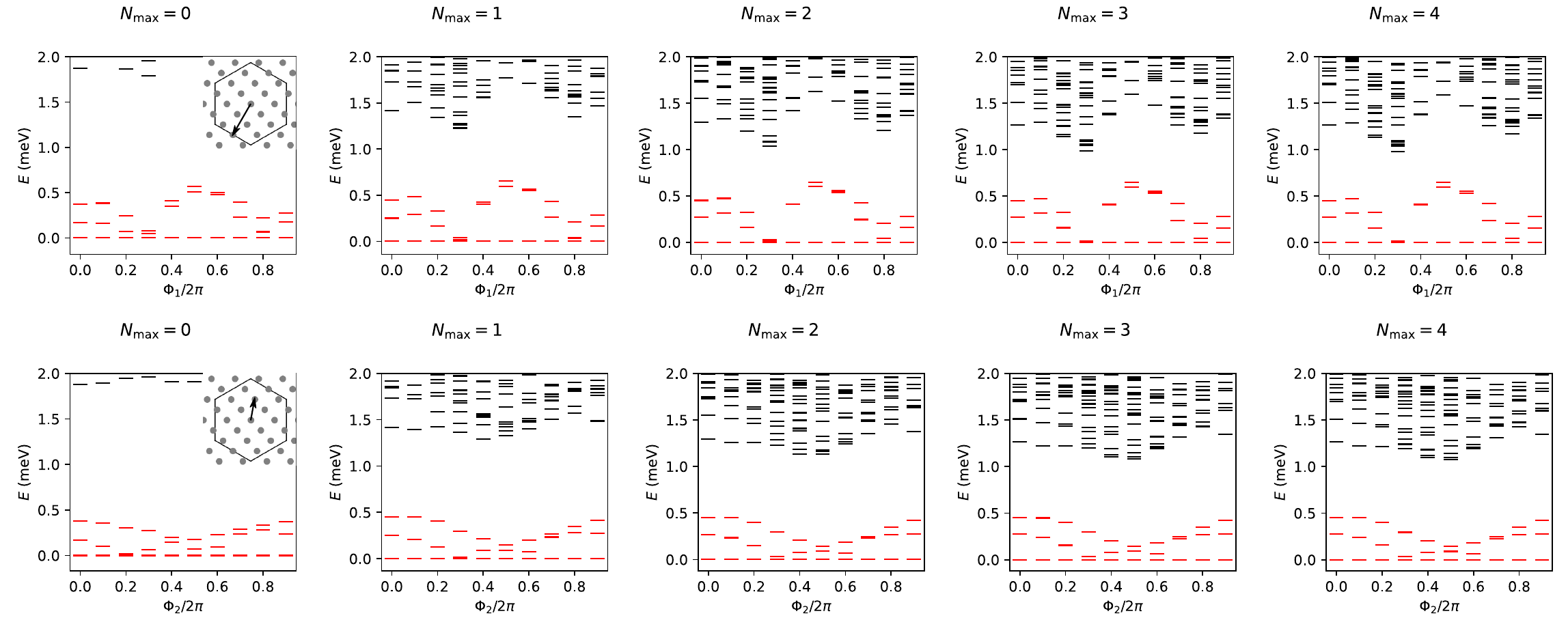}
    \caption{ 2-band $\nu=3+\frac{1}{3}$ ED calculations with bandmax restriction on the tilted $18$-site lattice for $\theta=1.76^\circ$, $\epsilon_r=8$ and $U=0\,$meV. The horizontal axis of each panel refers to the flux threaded along one handle of the torus. The columns show the many-body spectrum for different values of $N_\text{max}$, which denotes the maximum number of holes allowed in the lower HF band. The two rows correspond to threading flux along the two directions of the torus (see insets of the leftmost column for an illustration of the momentum mesh and the flux-threading direction). For the FCI momenta, we compute the two lowest states, while we only compute the lowest state for the non-FCI momenta. The lowest state in each FCI momenta is plotted in red. The many-body energies for each calculation are plotted relative to the lowest energy state across all momentum sectors.}
    \label{figapp:2x9_bandmax_flux}
\end{figure*}

In Fig.~\ref{figapp:2x9_bandmax_flux}, we present many-body spectra from 2-band ED calculations with bandmax restriction on a tilted $N=18$ lattice for $\theta=1.76^\circ$, $\epsilon_r=8$ and $U=0\,$meV, as a function of flux-threading along the two handles of the torus. For $N_\text{max}=0$, i.e.~a 1-band calculation, the system exhibits a three-fold quasi-degenerate ground state with the correct FCI momenta~\cite{regnaultFractionalChernInsulator2011}, and maintains a finite gap to the higher states under flux-threading. As $N_\text{max}$ is increased, the higher states drop in energy. However, the spectra are already well-converged by $N_\text{max}=2$. In particular, the three quasi-degenerate ground states remain well separated from the higher excitations, implying that the FCI survives in the 2-band limit. We caution that since the system size $N$ is still quite small, it is not possible to reliably extract quantitative phase diagrams for the FCI, but the calculations here demonstrate that FCIs can exist even when accounting for the full multi-band freedom of the central bands.

\end{appendix}

\end{document}